 \documentclass[final,3p,times]{elsarticle}

\usepackage{amssymb}
\usepackage{subfig}
\usepackage{amsmath}
\usepackage{mathrsfs}
\biboptions{sort&compress}
\DeclareMathAlphabet{\pazocal}{OMS}{zplm}{m}{n}

\newcommand{\Lb}{\pazocal{L}}

\usepackage[colorlinks=true,linkcolor=blue, citecolor=blue, urlcolor=blue]{hyperref}
\DeclareMathOperator{\sech}{sech}
\DeclareMathOperator{\sinc}{sinc}
\journal{Nuclear Physics B}

\begin{document}

\begin{frontmatter}



\title{Quasinormal modes of spherically symmetric black hole with cosmological constant and global monopole in bumblebee gravity }
 \author{Yenshembam Priyobarta Singh\fnref{label2}}
 \ead{priyoyensh@gmail.com}
 \author{Irengbam Roshila Devi\fnref{label2}}
 \ead{roshilairengbam@gmail.com}
 \author{Telem Ibungochouba Singh \corref{cor1}\fnref{label2}}
 \ead{ibungochouba@rediffmail.com}
 \affiliation[label2]{organization={Department of Mathematics, Manipur University},
            addressline={Canchipur}, 
            city={Imphal},
            postcode={795003}, 
            state={Manipur},
            country={India}}

%


\begin{abstract}
 This paper studies the scalar, electromagnetic field  and Dirac field perturbations of spherically symmetric black hole within the framework of Einstein-bumblebee gravity model with global monopole and cosmological constant. We investigate the effective potentials, greybody factors and quasinormal modes (QNMs) by applying the Klein-Gordon equation, electromagnetic field equation and Dirac equation expressed in Newman-Penrose (NP) formalism. Using the general method of rigorous bound, the greybody factors of scalar, electromagnetic and Dirac field are derived. Applying the sixth order WKB approximation and Pad\'{e} approximation, the QNM frequencies are derived. We also discuss the impact of global monopole $\eta$, cosmological constant $\Lambda$ and Lorentz violation parameter $L$ to the effective potential, greybody factor and QNMs. Increasing the parameter $\eta$ prevents the rise of effective potential for both Schwarzschild-de Sitter (SdS)-like and Schwarzschild-Anti de Sitter (SAdS)-like black holes with global monopole and consequently increases the greybody factors. However, decreasing the parameter $L$ reduces the rise of effective potential for the SAdS-like black hole with global monopole and it increases the greybody factor but increasing the parameter $L$ has an opposite effect for SdS-like black hole with global monopole. It is also shown that the shadow radius increases with increasing the parameter $\eta$ for both dS and AdS cases. Increasing the value of $L$ tends to increase the shadow radius for dS black hole but it has an opposite effect in AdS case. A careful studying is being carried out to investigate how the absorption cross-section gets affected when the parameter $L$ appears into the picture. 

\end{abstract}




\end{frontmatter}



\section{Introduction}
General Relativity is a theory which describes the gravitation at the classical level and is also plagued with the appearance of singularities. The standard model of particle physics has explained the other three fundamental interactions at the quantum front. A comprehensive understanding of the natural world can be obtained by using these two theories. Therefore, the unification of these two models is fundamental and its findings give us a better comprehension of nature. 
The perturbations of black hole theory plays a very important role in the exploration of quasinormal modes (QNMs). Ref. \cite{Label 1} discussed an innovative work on the black hole perturbation and the outcome of this study will provide a foundation for a plethora of other important research works. If a black hole is subjected to non radial perturbation, it oscillates with complex frequency at the intermediate stage. This oscillation is known as QNMs. The real part of QNMs is the oscillation frequency of the perturbation and the imaginary part is connected to damping \cite{S. L. D}. Ref. \cite{Label 2} discussed the QNMs in the simulation of gravitational wave scattering of a Schwarzschild space time. The QNMs represent the oscillation frequencies which heavily rely on the black hole  parameters like  mass, charge and spin \cite{Label 3,Label 5}. The results derived from the electromagnetic spectra  \cite{Label 6,Label 7} combined with gravitational waves will be able to investigate the fundamental parameters like mass, angular momentum and charge of the black hole, which in turn may be able to test the theory of gravity at the strong field limit. Since the observation of gravitational waves by LIGO and Virgo, a lot of interest has been paid in the study of QNMs. Many methods have been developed for calculating the QNMs such as WKB method \cite{B. F. S, S. I P, S. I C, W.}, analytical method \cite{S.C, A.O}, Frobenius method \cite{Label 4}, continued fraction method \cite{E.W.}, Mashhoon method \cite{H.-J}. Hawking \cite{Hawking1975} showed that the black holes emit particles as well as absorb, scatter and radiate when quantum gravity effect is taken into consideration. This emission is later known as Hawking radiation that interacts with the curvature of spacetime around the black hole event horizon when it leaves from the event horizon, accordingly altering its properties. An observer located at infinity cannot detect the usual black body spectrum \cite{Singleton2011,Cvetic1997}. A measure of variation between the black body spectrum and the rescaled spectrum is defined by the greybody factor. The greybody factor also represents a transmission coefficient in the scattering of black hole when the spacetime behaves the role of a potential barrier.  There are several methods for calculating
the greybody factors such as matching technique \cite{fernando2005,kim2008},
WKB approximation method for high gravitational potential \cite{parikh2000,konoplya2020a} and rigorous bound method \cite{visser1999,boonserm2008,boonserm2021,boonserm2014}.\\  There exists a great deal of uncertainty to determine the fundamental black hole parameters precisely, which gives us a lot of space for developing another modified gravity theory \cite{Label 8}.
The bumblebee gravity along with the other modified theories will be of particular interest, in which the Lorentz symmetry violation involves a non zero expectation value of a bumblebee field, if a suitable potential is taken into consideration. The Lorentz symmetry gives a fundamental symmetry for the derivation of any physically possible theories. The development of standard model of particle physics and general relativity are based on the Lorentz symmetry. These two theories have explained the known physical occurrences in the universe within the achievable energy range. 
The recent advances in unified gauge theories and the confirmation from high energy cosmic ray  \cite{Label 9,Label 10} proposed that Lorentz symmetry, which is taken as fundamental to quantum gravity theory and general relativity, break in physics at the high energy Planck scale $10^{19} $ GeV. Recently, some research works showed that the signal connected to Lorentz symmetry violation may precisely exhibit at lower energy level, which led to the discovery of their corresponding consequence in experiment \cite{Label 11}. Some observations which break the Lorentz symmetry at low energy residual effects of Planck scale are Cosmic rays \cite{Label 12, Label 13}, Photon tests \cite{Label 14, Label 15}, Muon tests \cite{Label 16, Label 17}, Meson tests \cite{Label 18, Label 19, Label 20}, Clock comparison experiments \cite{Label 21, Label 22}. Moreover, the Lorentz symmetry violation in the field of string theory, electrodynamics and non-abelian theory have been discussed in \cite{Label 23, Label 24, Label 25}. Since then, many researchers have studied the Lorentz symmetry violation in different types of spacetimes \cite{ Media2024, Onika2022, Priyo2022,  Media2023, Priyo2024, Onika2023}.

One of the important theories which includes the Lorentz symmetry violation is known as the Einstein-bumblebee gravity  \cite{V.A. K, R.V. M}. The prototype of the bumblebee model, a string inspired framework featuring tensor-induced Lorentz symmetry violation has been discussed in \cite{Label 27, V.A. K}. A potential $V(B^a B_a)$ acting on a vector field $B_a$ gives the spontaneous Lorentz symmetry violation in the frame of Bumblebee gravity \cite{Label 30}. Refs. \cite{Label 31, Label 32, Label 33, Label 34, Label 35, Label 36, Label 37, Label 38} investigated the properties of bumblebee model in Riemann and Riemann-Cartan spacetimes. An exact Schwarzschild-like black hole in the bumblebee gravity model has been derived by Casana et al. \cite{Label 39}. Since then, many researchers have investigated the bumblebee gravity model in different types of spacetimes such as traversable wormhole solution \cite{Label 40} , slowly rotating Kerr-like black holes \cite{Label 41}, Schwarzschild-like black holes with cosmological constant \cite{R.V. M} and other solutions  \cite{Label 43,Label 44,Label 45,Label 46,Label 47, Priyo24,Uniyal2023,Chen2023, Lin2023}. \\
    Furthermore, data which involve the Lorentz violation theory have been derived and were also retained in the black hole shadow  \cite{Label 41,H. W}, the black hole superradiance \cite{R. Jiang}, the accretion disc \cite{C. Liu} and the motion of massive body \cite{Label 33}. The use of quasi-periodic oscillation frequencies obtained from the observation data (GRO J1655-40, XTE J1550-564) and GRS1915+105 showed that the range of Lorentz spontaneous symmetry breaking becomes constrained for the rotating black hole within the bumblebee gravity model \cite{S.E. M, J.A. O, M.J. R}. 
 It is noted from the Grand Unified Theories that the global monopole is a special class of topological defect which may be developed in the early universe after the spontaneous breaking of symmetry of the global O(3) symmetry to U(1) \cite{T.W.B.}. The unusual property possessed by global monopole is a solid deficit angle, which makes the black hole with a global monopole topologically different and also leads to an interesting outcome in the physical consequence \cite{Q. Pan, S. Chen}. Ref. \cite{gullu2022} derived an exact Schwarzschild-like black hole solution with global
monopole in bumblebee gravity and studied the effect of global monopole and bumblebee field on photon sphere and black hole shadow.\\

The shadow cast by a Schwarzschild black hole is initially investigated by Synge \cite{J.L.}. The influence of a thin accretion disc on the shadow of black hole is investigated by Luminet \cite{J.P.}. The shadow of rotating Kerr black hole is also studied by Bardeen \cite{J.M.}. After taking the direct image of M8$7^*$ at the core of the Virgo A galaxy and Sgr$A^*$ at the core of the Milky Way Galaxy \cite{K. A, V.I.}, many researchers have paid a lot of interest in the study of black hole shadow. The internal structures of the black hole cannot be studied directly due to their special properties. The black hole interacts with the surrounding environment, such as scattering, absorption and Hawking radiation. This gives the information about the internal structure of the black hole. To study the absorption cross-section as one of the interactions is very important. The most efficient and useful way to understand the black hole properties is to study the matter wave absorption and test the field around black holes. Since then many researchers \cite{Higuchi2001, Kanti2002, Jung2004, Grain2005, L.C.B.2007, L.C.B.2009, C.F.B.2014, Huang2015, L.C.S.2018, Huang2019, Anacleto2020, Magalhães2020, H.C.D.2020, Benone2018, Li2022} studied the absorption cross-section in different black hole spacetimes. In this connection, the paper aims to investigate the effects of monopole parameter $\eta$ and Lorentz violation parameter $L$ in the greybody factor, QNMs, shadow radius and absorption cross-section of spherically symmetric black hole.   \\

The paper is organised as follows: In Section 2, we discuss come properties of SdS/AdS-like black hole with global monopole. We discuss the perturbation of scalar, electromagnetic field and Dirac field and the corresponding effective potentials of the black hole are also derived in Section 3. In section 4, the greybody factors from the scalar, electromagnetic and Dirac fields are derived. In section 5, the quasinormal mode frequencies for scalar, electromagnetic and Dirac perturbations have been derived by using WKB 6th order and Pad\'e approximation for the different black hole parameters. We discuss the shadow radius in Section 6 and high energy absorption cross section in Section 7. Conclusion is given in Section 8.

\section{Schwarzschild-like black hole with global monopole and \textbf{cosmological constant} }    
The line element of Schwarzschild-like black hole with global monopole and cosmological constant in bumblebee gravity is given by \cite{Gogoi2022}
\begin{eqnarray} \label{eqn line}
ds^2&=&f(r)dt^2-\left(\dfrac{1+L}{f(r)}\right)dr^2- r^2 d\theta^2 - r^2 \sin^2\theta d\varphi^2,
\end{eqnarray}
where  $ f(r) $ has the form
\begin{eqnarray}\label{eqn f}
f(r)=1-\kappa\eta^2 -\dfrac{2M}{r}-\dfrac{ (1+L)\Lambda r^2}{3},
\end{eqnarray}
where $M$ is the mass of the black hole, $L$ is the Lorentz violating parameter, $\kappa=8\pi$, $\Lambda$ is the cosmological constant and $\eta$ is the global monopole parameter. Clearly in the limiting value of $L\rightarrow 0$, Eq. \eqref{eqn line} reduces to the line element of SdS/AdS black hole with global monopole. If $\eta=0$, the resulting metric reduces to the line element of SdS/AdS-like black hole in bumblebee gravity \cite{R.V. M}. Moreover if $\Lambda = 0$, we recover the line element
of Schwarzschild-like black hole with global monopole \cite{gullu2022}. The expression for the black hole mass $M$ is obtained from $f(r)=0$ as
\begin{align}\label{eqn mass}
M=\dfrac{r}{6} \left\lbrace 3 \left(1-\kappa \eta^2 \right)- (1+L) r^2 \Lambda \right\rbrace.
\end{align}

It is noticed that $f(r)\rightarrow -\infty$ as $r\rightarrow 0$. Further, as $r\rightarrow \infty$, $f(r) \rightarrow \pm \infty$ according to the positive or negative cosmological constant provided. For AdS case, $f(r)$ keeps on increasing but for dS case i.e. $\Lambda>0$, the value of  $f(r)$ increases at first  and then decreases. It is to be noted that there is only one positive real horizon for AdS case. However, for dS case, in order to get two positive real horizons, the maximum value of $f(r)$ must be positive. The maximum point of $f(r)$ is achieved at $r=r_m= \left\lbrace\frac{3M}{\Lambda (1+L)}\right\rbrace^{1/3}$. The condition for having two positive real horizons for dS case is obtained as
\begin{align}
\left(1-\kappa \eta^2 \right)^3 > 9 M^2 \Lambda (1+L).
\end{align}
The event horizon $r_e$ and the cosmological horizon $r_c$ for dS spacetime are given by
\begin{eqnarray}
r_e=2\sqrt{\dfrac{1-\kappa \eta^2}{\Lambda (1+L)}}\cos \left( \dfrac{\pi +\Phi}{3}\right) ,\\
r_c=2\sqrt{\dfrac{1-\kappa \eta^2}{\Lambda (1+L)}}\cos \left( \dfrac{\pi -\Phi}{3}\right) ,
\end{eqnarray}
where $ \Phi =\cos^{-1}\left[\dfrac{3M\sqrt{\Lambda (1+L)}}{\sqrt{(1-\kappa \eta ^{2})^3}} \right]  $ .


\begin{figure}[h!]
\centering
  \subfloat[\centering ]{{\includegraphics[width=170pt,height=170pt]{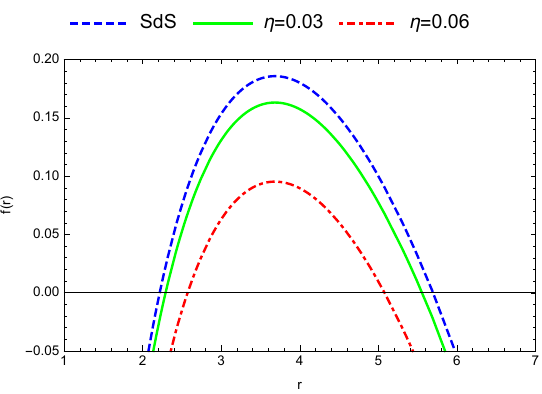}}\label{fig fpo}}
  \qquad
   \subfloat[\centering ]{{\includegraphics[width=170pt,height=170pt]{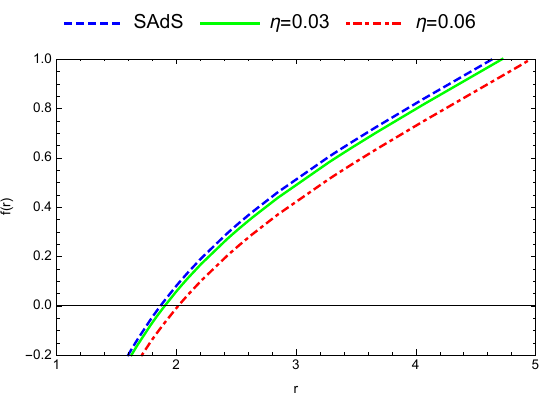}}\label{fig fle}}
   \caption{Variation of the metric function of the Schwarzschild-dS/AdS with global monopole under bumblebee gravity model. The physical parameters are chosen as (a) $M=1$, $\Lambda=0.05$ and (b) $M=1$, $\Lambda=-0.05$.}
   \label{fig f}
\end{figure}

\begin{figure}[h!]
\centering
  \subfloat[\centering ]{{\includegraphics[width=170pt,height=170pt]{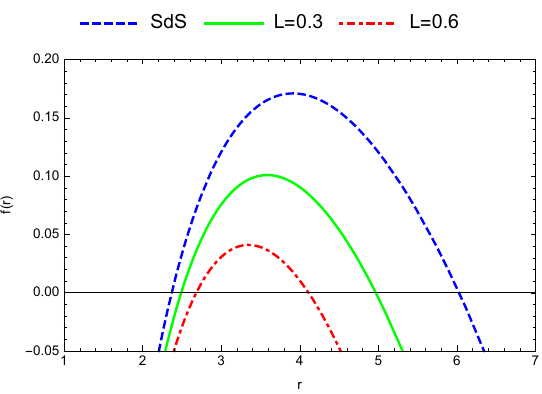}}\label{fig fLpo}}
  \qquad
   \subfloat[\centering ]{{\includegraphics[width=170pt,height=170pt]{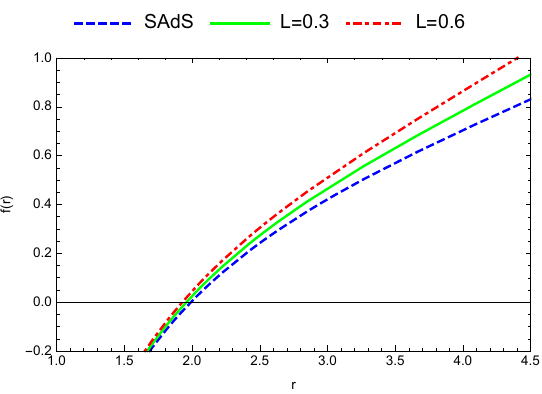}}\label{fig fLne}}
   \caption{Variation of the metric function of the Schwarzschild-dS/AdS with global monopole under bumblebee gravity model. The physical parameters are chosen as (a) $M=1$, $\Lambda=0.05$ and (b) $M=1$, $\Lambda=-0.05$.}
   \label{fig fL}
\end{figure}
Figs. \ref{fig f} and \ref{fig fL} illustrate the behaviour of metric function for both dS and AdS spacetimes for different values of monopole parameter and Lorentz violation parameter  respectively. For SAdS-like black hole, there is only one horizon (event) and the horizon radius increases with increasing the global monopole parameter and Lorentz violation parameter. Further, it is evident from Fig. \ref{fig f} that for SdS-like black hole, there are two horizons: event horizon and cosmological horizon. Increasing the global monopole parameter and Lorentz violation parameter increase the distance between the horizons.


\section{Scalar, electromagnetic and Dirac perturbations}
Black hole perturbation is a useful tool to study small disturbances in black hole spacetime. The study of black hole perturbation is important in understanding the black hole stability, the emission of gravitational waves, Hawking evaporation and the intricate dynamics of extreme astrophysical environments. Black hole perturbation arises due to scalar, electromagnetic, gravitational or Dirac field. When black hole interacts with its surrounding it  undergoes perturbation thereby emitting gravitational waves and the waves are detected by observatories like LIGO and Virgo. The study of perturbations of black hole under Bumblebee gravity model has gained interest among researchers to explore the  modifications to the gravitational wave propagation and quasinormal mode. Different types of perturbation for different black holes are widely studied in \cite{Priyo24,Devi2020, Gogoi2023,Lin2023, Gogoi2022,Uniyal2023,Chen2023}. In our work, we will study the perturbation of each field separately  by assuming the   perturbing  one  field  does  not  affect  the   background of the other fields.

\subsection{Scalar perturbation}
The evolution of the scalar field is governed by the Klein-Gordon equation, which is given by \cite{Ibungo, Ibungo2015}
\begin{eqnarray}\label{eqn klein}
\dfrac{1}{\sqrt{-g}}\partial_\mu \left(\sqrt{-g}~g^{\mu \nu}     \partial_\nu \Psi \right)=0,
\end{eqnarray}
where  $g^{\mu \nu}$ and $g$ are the inverse  of the metric tensor and determinant of the metric tensor. Since the line element \eqref{eqn line} represents a static and spherically symmetric metric, the radial and angular parts can be separated by choosing 
\begin{align}
\Psi(t,r,\theta,\phi)= e^{-i\omega t}R(r) ~Y_{lm}(\theta,\phi), 
\end{align}
where $ Y_{lm}(\theta,\phi)$ are the standard scalar spherical harmonics.

The radial part of Eq. \eqref{eqn klein}, reduces to a one-dimensional Schr\"{o}dinger-like equation as 
\begin{eqnarray}
\dfrac{d^{2}\Psi_h}{dr_*^{2}}+\left(\omega^{2}-V_s(r)\right)\Psi_h=0,
\end{eqnarray}
where $r_*$ is the tortoise coordinate defined by
\begin{align}
dr_*=\dfrac{\sqrt{1+L}}{f(r)}dr
\end{align}
and $V_s(r)$ is the effective potential for the perturbation due to scalar field which is derived as
\begin{align}\label{eqn Vs}
V_{s} (r)= f(r) \left\lbrace \dfrac{l(l+1)}{r^2}+\dfrac{2M}{(1+L)r^3}-\dfrac{2\Lambda}{3} \right\rbrace.
\end{align}

\begin{figure}[h!]
\centering
  \subfloat[\centering ]{{\includegraphics[width=170pt,height=170pt]{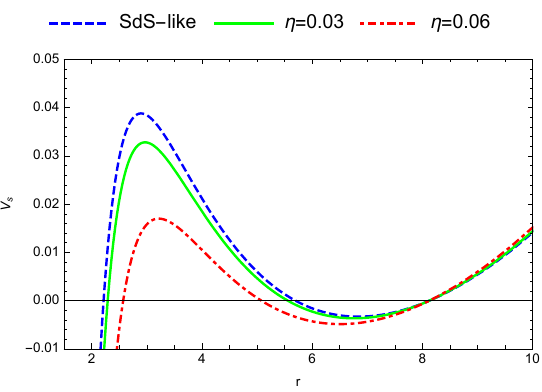}}\label{fig SVepo}}
  \qquad
   \subfloat[\centering ]{{\includegraphics[width=170pt,height=170pt]{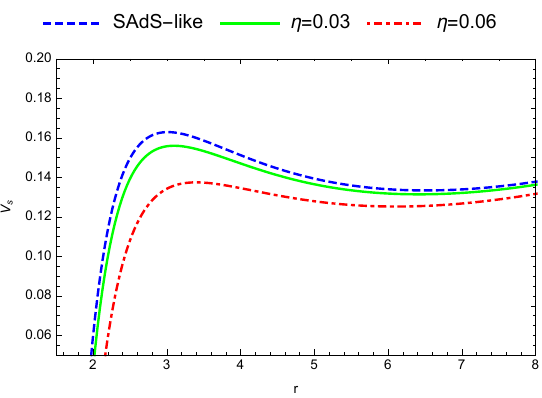}}\label{fig SVene}}
   \caption{Graphs of the effective potential of the scalar field perturbation: (a) SdS-like black hole ($\Lambda=0.05$) and (b) SAdS-like black hole ($\Lambda=-0.05$) for different values of  global monopole. The physical parameters are chosen as  $M=1$, $l=1$ and $L=0.2$.}
   \label{fig SVe}
\end{figure}

Fig. \ref{fig SVe} shows the variation of effective potential of scalar field perturbation for different values of monopole parameter. Here blue dashed line represents the effective potential of scalar field perturbation for SdS-like/SAdS-like black hole in the absence of monopole parameter. For dS case,  it is observed from Fig. \ref{fig SVepo} that the effective potential vanishes at the event horizon $(r_e)$, cosmological horizon $(r_c)$ and another additional point $r_a$ respectively. The vanishing of the effective potential at the horizons is clearly evident since $f(r)$ appears as a factor in the effective potential. The additional zero point $r_a$ is the point where the second factor in the RHS Eq. \eqref{eqn Vs} becomes zero. Since $\eta$ doesn't appear in the second factor, the third vanishing point ($r_a$) remains the same for all the values of $\eta$ (see Fig. \ref{fig SVepo}). For AdS case, $\Lambda$ is negative, so the second factor of the effective potential is strictly positive. Therefore the vanishing point of the effective potential is only at the point where $f(r)$ vanishes i.e. at event horizon. For both dS and AdS cases, the local maximum of the effective potential tends to decrease as the monopole parameter increases. Further, the effective potential of dS case acts as a decreasing function of $\eta$ in the range $r_e<r<r_a$ but in the range $r_a<r$, it acts as an increasing function of $\eta$ . However, for AdS case the effective potential behaves as a decreasing function of $\eta$. To discuss the results analytically, we take the derivative of $V_s$ with respect to $\eta$ as
\begin{align}\label{eqn dVse}
\dfrac{d V_{s}}{d \eta}
=& -2 \kappa \eta ~ \Sigma_1 ,
\end{align}
where $\Sigma_1= \dfrac{l(l+1)}{r^2}+\dfrac{2M}{(1+L)r^3}-\dfrac{2\Lambda}{3} $. For dS case, one can clearly see that $V_s$ is a decreasing or increasing function  of $\eta$ according to $\Sigma_1>0$ or $\Sigma_1<0$. Therefore, in the region $r_e<r<r_a$, $V_s$ is a decreasing function  of $\eta$ but  in the region $r_a<r$, it is an increasing function. However, for AdS case, $V_s$ is a decreasing function of $\eta$ in the entire range since $\Sigma_1>0$. The findings are consistent with the graphical results obtained in Fig. \ref{fig SVe}.

\begin{figure}[h!]
\centering
  \subfloat[\centering ]{{\includegraphics[width=170pt,height=170pt]{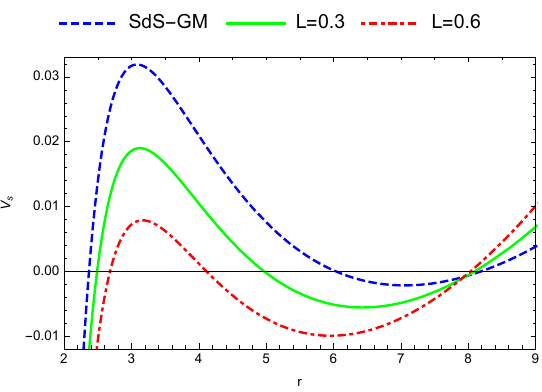}}\label{fig SVLLpo}}
  \qquad
   \subfloat[\centering ]{{\includegraphics[width=170pt,height=170pt]{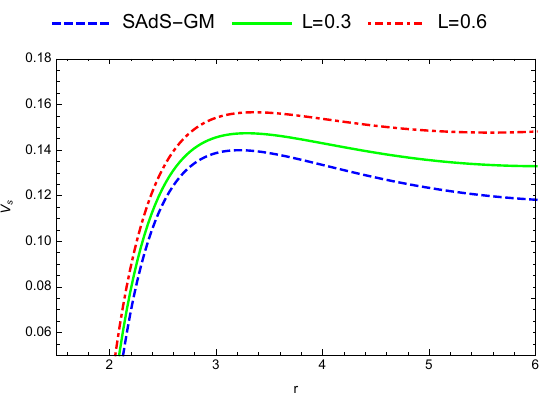}}\label{fig SVLLne}}
   \caption{Graphs of the effective potential of the scalar field perturbation: (a) SdS-like black hole with global monopole ($\Lambda=0.05$) and (b) SAdS-like black hole with global monopole ($\Lambda=-0.05$) for different values of  Lorentz violation parameter $L$. The physical parameters are chosen as  $M=1$, $l=1$ and $\eta=0.05$.}
   \label{fig SVLL}
\end{figure}

In order to see the effect of bumblebee gravity on the effective potential, we illustrate the behaviour of effective potential  in Fig. \ref{fig SVLL}  by varying Lorentz violation parameter and fixing other parameters. The blue dashed line in Figs. \ref{fig SVLLpo} and \ref{fig SVLLne} represent the original SdS and SAdS black hole with global monopole. For dS case, the local maximum of the effective potential decreases with increasing the Lorentz violation parameter but for AdS case the effect is opposite. The effective potential $V_s$ decreases with increasing $L$ in between the event horizon and the additional zero point $r_a$ for the dS case. However, it behaves as  an increasing function of $L$ when $r>r_a$. Therefore, in the region between the event horizon and the cosmological horizon, when the Lorentz violation parameter increases in the dS case, less wave energy is being communicated to the surrounding medium which is an  indication of higher greybody factors. Conversely, in the case of AdS, raising the Lorentz violation parameter traps more wave energy, which lowers the greybody factors. To examine the above results analytically we calculate
\begin{align}\label{eqn dVsL}
\dfrac{d V_s}{d\,L}= \Sigma_2=\dfrac{4M^2}{(1+L)^2 r^4} +\dfrac{2r^2 \Lambda^2}{9}-\dfrac{2\left(1-\kappa \eta^2 \right)}{(1+L)^2 r^3}-\dfrac{l(1+l)\Lambda}{3}.
\end{align}

From Eq. \eqref{eqn dVsL}, one can clearly see that for both dS and AdS cases, $\Sigma_2$ can be positive or negative according to the choice of the values of black hole parameters. Thus,  according to the range of r and the choice of the black hole parameter, the effective potential may  increase or decrease with increasing $L$ which is consistent with Fig.  \ref{fig SVLLpo}. 

\subsection{Electromagnetic perturbation}
In this subsection, we will study the electromagnetic field perturbation of Schwarzschild dS/AdS black hole with global monopole in the bumblebee gravity. 

The electromagnetic field in curved spacetime follows the equation \cite{Zhang2021}
\begin{eqnarray}\label{eqn em}
&\dfrac{1}{\sqrt{-g}}\partial_\nu \left(\sqrt{-g}~g^{\mu \alpha}g^{\nu \beta}F_{\alpha \beta} \right)=0,
\end{eqnarray}
where the Maxwell tensor is $ F_{\mu \nu}=\partial_\mu A_\nu -\partial_\nu A_\mu $ and $A_\mu$ is a four dimensional vector potential.
In the Regge-Wheeler-Zerilli formalism, one may decompose $A_\mu$ in terms of the scalar  and vector  spherical harmonics
\begin{eqnarray}\label{eqn au}
A_\mu=\sum_{l,m}e^{-i\omega t}\left(\begin{bmatrix}
0\\o\\a^{lm}(r)\textbf{S}_{lm}\\
\end{bmatrix}+ \begin{bmatrix}
j^{lm}(r)Y_{lm}\\h^{lm}(r)Y_{lm}\\k^{lm}(r)\textbf{Y}_{lm}\\
\end{bmatrix}\right),
\end{eqnarray} where $Y_{lm}$ are the scalar spherical harmonics and   the vector spherical harmonics ($\textbf{S}_{lm}$ and $\textbf{Y}_{lm}$) are defined as 
\begin{eqnarray}
\textbf{S}_{lm}=\begin{pmatrix}\dfrac{1}{\sin \theta}\partial_\varphi Y_{lm}\\-\sin \theta ~ \partial_\theta Y_{lm} 
\end{pmatrix},\hspace{0.4cm} \textbf{Y}_{lm}= \begin{pmatrix}\partial_\theta Y_{lm}\\ \partial_\varphi Y_{lm} 
\end{pmatrix},
\end{eqnarray}
where $\omega$ is the frequency. Here $l$ and $m$ denote the angular momentum quantum number and the azimuthal number respectively. The first term in the right hand side of Eq. \eqref{eqn au} represents the axial mode while the second term represents the polar mode. It's important to note that the axial and polar modes have parity $(-1)^{l+1}$ and $(-1)^l$ respectively. Moreover, the polar and axial parts have equal contributions  to  the  final  result  \cite{wheeler1973,ruffini1973}. Thus, we focus only on the axial part. Now, substituting Eq. \eqref{eqn au} in Eq. \eqref{eqn em} and applying tortoise coordinate $r_*$, the radial part  of Eq. \eqref{eqn em} can be  written in the form of Schr\"{o}dinger like wave form as

\begin{align}
\dfrac{d^{2} a^{lm}}{dr_*^{2}}+\left(\omega^{2}-V_h(r)\right)a^{lm}=0,
\end{align} 

where 
\begin{align}\label{eqn Vm}
V_{m}(r)=f(r) \times \dfrac{l(l+1)}{r^2}.
\end{align}

\begin{figure}[h!]
\centering
  \subfloat[\centering dS ]{{\includegraphics[width=170pt,height=170pt]{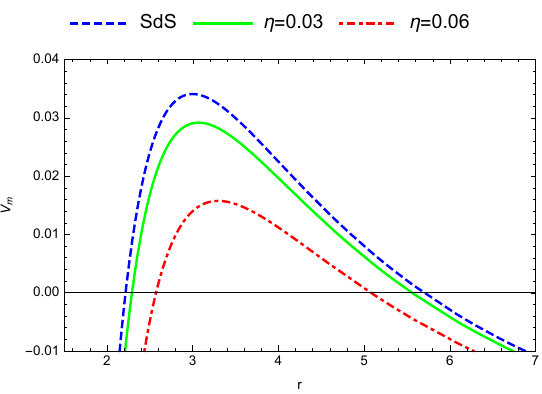}}\label{fig Vmepo}}
  \qquad
   \subfloat[\centering  AdS]{{\includegraphics[width=170pt,height=170pt]{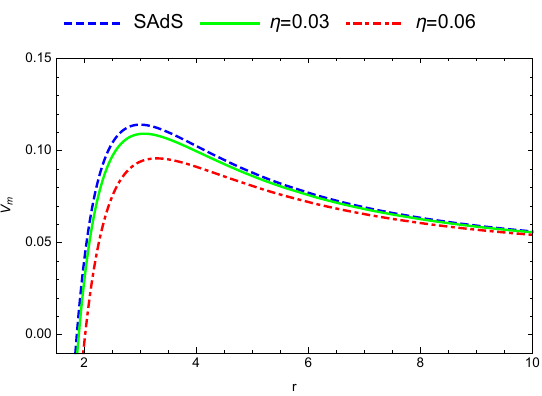}}\label{fig Vmene}}
   \caption{Graphs of the effective potential of the electromagnetic field perturbation: (a) SdS-like black hole ($\Lambda=0.05$) and (b) SAdS-like black hole ($\Lambda=-0.05$) for different values of  global monopole. The physical parameters are chosen as  $M=1$, $l=1$ and $L=0.2$.}
   \label{fig Vme}
\end{figure}


To analyse the effect of $\eta$ on the effective potential of electromagnetic field perturbation, we plot $V_m$ of dS and AdS cases for different values of $\eta$ in Figs. \ref{fig Vmepo} and \ref{fig Vmene} respectively. Both the effective potentials of dS and AdS vanish at the horizons. In contrast to the effective potential of scalar perturbation, which vanishes at three points, the electromagnetic perturbation's effective potential vanishes at only two points in the dS case. Increasing the values of monopole parameter the effective potential reduces in both dS and AdS cases. Thus, the monopole parameter $\eta$ enhances the flow of waves to the surrounding in the electromagnetic field perturbation. This suggests the greybody factor of the electromagnetic perturbation will get higher with increasing the monopole parameter. Further, we will also analyse the nature of the effective potential analytically. From Eq. \eqref{eqn Vm}, we calculate
\begin{align}\label{eqn dVm}
\dfrac{dV_m}{d\eta}=\dfrac{-2\kappa \eta l (1+l)}{r^2}.
\end{align}
From Eq. \eqref{eqn dVm}, it is clearly evident that for both the dS and AdS cases, the effective potential $V_m$ is a decreasing function of $\eta$.

\begin{figure}[h!]
\centering
  \subfloat[\centering ]{{\includegraphics[width=170pt,height=170pt]{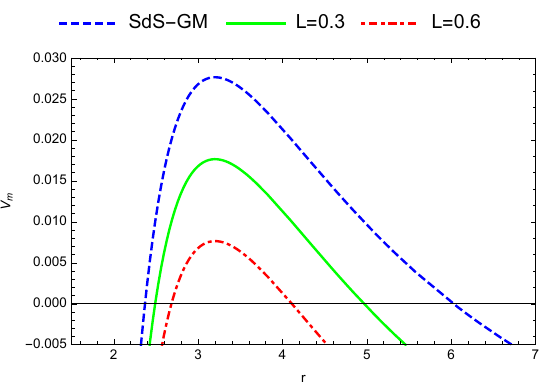}}\label{fig VmLpo}}
  \qquad
   \subfloat[\centering ]{{\includegraphics[width=170pt,height=170pt]{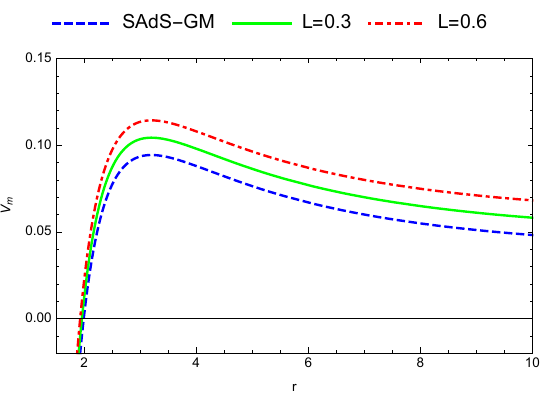}}\label{fig VmLne}}
   \caption{Graphs of the effective potential of the electromagnetic field perturbation: (a) SdS-like black hole ($\Lambda=0.05$) and (b) SAdS-like black hole ($\Lambda=-0.05$) for different values of  Lorentz violation parameter. The physical parameters are chosen as  $M=1$, $l=1$ and $\eta=0.05$.}
   \label{fig VmL}
\end{figure}

The nature of the effective potential of electromagnetic perturbation for dS and AdS for different values of $L$ are illustrated in Figs. \ref{fig VmLpo} and \ref{fig VmLne} respectively. For dS case, the effective potential decreases with increasing the Lorentz violation parameter. Since the potential barrier reduces as $L$ increases, more waves are transmitted, thereby indicating the rise of greybody factor. However for AdS case, the potential barrier increases with increasing the Lorentz violation parameter $L$. Thus the effect of Lorentz violation parameter on the effective potential of electromagnetic perturbation of dS and AdS cases are completely opposite. This opposite effect can be also observed from the analytical analysis of $V_m$. The derivative of $V_m$ with respect to $L$ can be derived from Eq. \eqref{eqn Vm} as

\begin{align}
\dfrac{dV_m}{dL}=- \dfrac{l(1+l) \Lambda}{3}.
\end{align}
We observe that according to the sign of $\Lambda$, $dV_m/dL$ can be either positive or negative. Thus, $V_m$ is a decreasing function in dS case but it is an increasing function in AdS case.

\subsection{Dirac perturbation}
In this section, we will discuss the massless and massive Dirac perturbation of SdS/SAdS-like black hole in global monopole using the NP formalism.  In NP formalism, the Chandrasekhar-Dirac equation (CDE) is given by \cite{Chandrashekhar}
\begin{eqnarray} \label{eqn CDE}
&(D+\epsilon-\rho)B_1 + (\overline{\delta}+\pi-\alpha)B_2=i\mu^*C_1, \nonumber \\
&(\delta+\beta-\tau)B_1+(\Delta+\mu-\gamma)B_2=i\mu^*C_2,\nonumber\\
&(D+\overline{\epsilon}-\overline{\rho})C_2-(\delta+\overline{\pi}-\overline{\alpha})C_1=i\mu^*B_2, \nonumber \\
&(\Delta+\overline{\mu}-\overline{\gamma})C_1- (\overline{\delta}+ \overline{\beta}-\overline{\tau})C_2=i\mu^*B_1,
\end{eqnarray}
where $B_1$, $B_2$, $C_1$ and $C_2$ represent the Dirac spinors and $D=l^\mu \partial_\mu$, $\Delta=n^\mu \partial_\mu$, $\delta=m^\mu \partial_\mu$ and $\bar{\delta}= \overline{m}^\mu \partial_\mu$ are the directional
derivatives. $\epsilon$, $\rho$, $\pi$, $\alpha$, $\beta$, $\tau$, $\mu$ and $\gamma$ are the spin  coefficients and  the bar over the quantity denotes complex conjugation. The  null tetrad basis vectors are choosen as

\begin{align}
& l^{\mu}=\left\lbrace \left( 1-\kappa \eta^2-\dfrac{2M}{r}-\dfrac{(1+L)\Lambda r^2}{3}\right)^{-1} ,\dfrac{1}{\sqrt{1+L}},0,0\right\rbrace,\nonumber \\
& n^{\mu}=\left\lbrace \dfrac{1}{2},\dfrac{-1}{2\sqrt{1+L}}\left(1-\kappa \eta^2-\dfrac{2M}{r}-\dfrac{(1+L)\Lambda r^2}{3}\right) ,0,0\right\rbrace,\nonumber\\ 
& m^{\mu}=\left\lbrace 0,0,\dfrac{1}{\sqrt{2}r},\dfrac{i\csc \theta}{r\sqrt{2}}\right\rbrace,\nonumber \\ 
& \overline{m}^{\mu}=\left\lbrace 0,0,\dfrac{1}{\sqrt{2}r},\dfrac{i\csc \theta}{r\sqrt{2}}\right\rbrace.
\end{align}
The non-zero spin  coefficients are found to be
\begin{eqnarray}\label{eqn spin}
&\rho = \dfrac{-1}{r \sqrt{1+L}}, \hspace{0.3cm} \mu =\dfrac{-1}{2 r\sqrt{1+L}} \left( 1-\kappa \eta^2-\dfrac{2M}{r}-\dfrac{(1+L)\Lambda r^2}{3}\right) , \nonumber \\
& \gamma =\dfrac{1}{4\sqrt{1+L}}\left(\dfrac{2M}{r^2}-\dfrac{2(1+L)\Lambda r}{3} \right) ,
\hspace{0.3cm} \beta = \dfrac{\cot \theta}{2\sqrt{2r}}, \hspace{0.3cm} \alpha= -\dfrac{\cot \theta}{2\sqrt{2r}}.
\end{eqnarray}
To decouple CDE \eqref{eqn CDE} into radial and angular parts, we consider the spin-1/2 wave function in the
form
\begin{eqnarray}\label{eqn spinor}
B_1=R_1(r)A_1(\theta)e^{[i(\omega t+\widetilde m \phi)]},\nonumber \\
C_1=R_2(r)A_1(\theta)e^{[i(\omega t+\widetilde m \phi)]},\nonumber \\
B_2=R_2(r)A_2(\theta)e^{[i(\omega t+\widetilde m \phi)]},\nonumber\\
C_2=R_1(r)A_2(\theta)e^{[i(\omega t+\widetilde m \phi)]},
\end{eqnarray} 
where $\omega$ is the frequency of the incoming Dirac field and $m$ is the azimuthal quantum number of the wave. Substituting the spin coefficients \eqref{eqn spin} and the spinors \eqref{eqn spinor} in Eq. \eqref{eqn CDE}, the radial and the angular parts are obtained as

\begin{align}\label{eqn radial}
&\left[ \dfrac{1}{\sqrt{1+L}}\left( 1+r\dfrac{d}{dr}\right)+i r \omega \left( 1-\kappa \eta^2-\dfrac{2M}{r}-\dfrac{(1+L)\Lambda r^2}{3}\right)^{-1} \right]R_1-i\mu^*r R_2  =-\lambda_1 R_2,\nonumber \\
& \left[\dfrac{1}{\sqrt{1+L}}\left(  1-\kappa \eta^2-\dfrac{2M}{r}-\dfrac{(1+L)\Lambda r^2}{3}\right) \left( 1+r\dfrac{d}{dr}\right) + \dfrac{r}{2\sqrt{1+L}}
\left(\dfrac{2M}{r^2}-\dfrac{2(1+L)\Lambda r}{3} \right) -i r \omega\right]R_2 \nonumber\\ & +2i\mu^*r R_1=\lambda_2 R_1,\nonumber\\
&\left[ \dfrac{1}{\sqrt{1+L}}\left( 1+r\dfrac{d}{dr}\right)+ i r \omega \left( 1-\kappa \eta^2-\dfrac{2M}{r}-\dfrac{(1+L)\Lambda r^2}{3}\right)^{-1}      
\right] R_1 -i\mu^*r R_2=\lambda_3 R_2,\nonumber \\
&  \left[ \dfrac{1}{\sqrt{1+L}}\left(  1-\kappa \eta^2-\dfrac{2M}{r}-\dfrac{(1+L)\Lambda r^2}{3}\right) \left( 1+r\dfrac{d}{dr}\right) + \dfrac{r}{2\sqrt{1+L}}
\left(\dfrac{2M}{r^2}-\dfrac{2(1+L)\Lambda r}{3} \right) -i r \omega     \right] R_2  \nonumber\\ &+2i\mu^*r R_1=-\lambda_4 R_1.
\end{align}
and 
\begin{align}\label{eqn angular}
&\widetilde L^+ A_2=\sqrt{2}A_1\lambda_1, \hspace{0.2cm} \widetilde L^{-} A_1=\dfrac{1}{\sqrt{2}}A_2\lambda_2,\nonumber \\
& \widetilde L^{-} A_1=\sqrt{2}A_2\lambda_3, \hspace{0.2cm} \widetilde L^+ A_2=\dfrac{1}{\sqrt{2}}A_1\lambda_4,
\end{align}
 where $\lambda_1$, $\lambda_2$, $\lambda_3$ and $\lambda_4$ are the separation constants. $ \widetilde L^{\pm}$  are the angular operators, defined as
\begin{align}
&\widetilde L^\pm=\dfrac{d}{d\theta}\pm \dfrac{\widetilde m}{\sin\theta}+\dfrac{\cot\theta}{2}.
\end{align}
We take the separation constant as 
\begin{align}
\sqrt{2}\lambda_1=\dfrac{1}{\sqrt{2}}\lambda_1=-\lambda  ~~\text{and} \hspace{0.2cm} \dfrac{1}{\sqrt{2}}\lambda_2=\sqrt{2}\lambda_3=\lambda.
\end{align} 
Setting the eigenvalue $\lambda=-(l+1/2)$ for the angular equation, one can have the spin-weighted spheroidal harmonics \cite{newman1967,goldberg1967}. We carry out the following transformation to get the radial equations in the form of one-dimensional Schr\"{o}dinger-like wave equations

\begin{align}
R_1(r)=\dfrac{P_+(r)}{r},  ~~~R_2(r)=\dfrac{1}{r} \left(1-\kappa \eta^2-\dfrac{2M}{r}-\dfrac{(1+L)\Lambda r^2}{3} \right)^{-\frac{1}{2}} P_-(r).
\end{align}
The radial equation becomes
\begin{align}\label{eqn dirac radial}
&\left( \dfrac{d}{dr_*}\pm i\omega\right) P_\pm =\sqrt{1-\kappa \eta^2-\dfrac{2M}{r}-\dfrac{(1+L)\Lambda r^2}{3}}\left( \dfrac{\lambda}{r}\pm i\mu_*\right) P_\mp ,
\end{align}
where $\mu_*$ is the normalized rest mass of the spin-1/2 particle and $r_*$ is the tortoise coordinate defined by
\begin{align}
\dfrac{d}{dr_*}=\dfrac{1}{\sqrt{1+L}}\left( 1-\kappa \eta^2-\dfrac{2M}{r}-\dfrac{(1+L)\Lambda r^2}{3}\right) \dfrac{d}{dr},
\end{align}
\subsubsection{Massless Dirac field perturbation}
For massless Dirac field perturbation $\mu_*=0$, and further letting 
\begin{align}
&Z_\pm=P_- \pm P_+,
\end{align}

we obtain a pair of one dimensional Schr\"{o}dinger-like wave equations
\begin{align}
\left( \dfrac{d^2}{dr_*^2}+\omega ^2\right) Z_\pm=V_{d \pm}Z_\pm, 
\end{align} 
where $V_{d \pm}$ are the effective potentials for the massless Dirac field :
\begin{align}
V_{d \pm}=& \dfrac{\left(l+\frac{1}{2}\right)^2}{r^2}\left( 1-\kappa \eta^2-\dfrac{2M}{r}-\dfrac{(1+L)\Lambda r^2}{3}\right) \nonumber\\ & \pm  \left(l+\frac{1}{2}\right)\sqrt{1-\kappa \eta^2-\dfrac{2M}{r}-\dfrac{(1+L)\Lambda r^2}{3}}  \times \dfrac{ \left\lbrace 3M-r(1-\kappa \eta ^2) \right\rbrace}{r^3 \sqrt{1+L}}.
\end{align}

\begin{figure}[h!]
\centering
  \subfloat[\centering  ]{{\includegraphics[width=170pt,height=170pt]{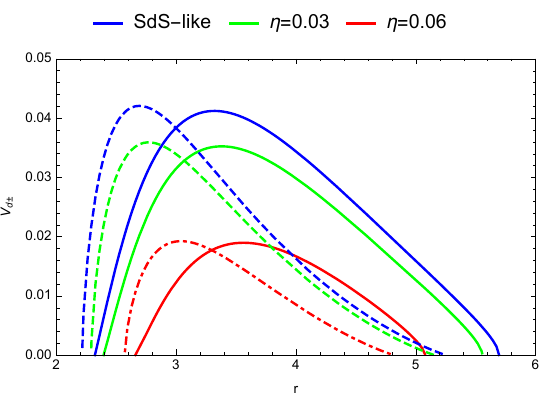}}\label{fig Vdepo}}
  \qquad
   \subfloat[\centering  ]{{\includegraphics[width=170pt,height=170pt]{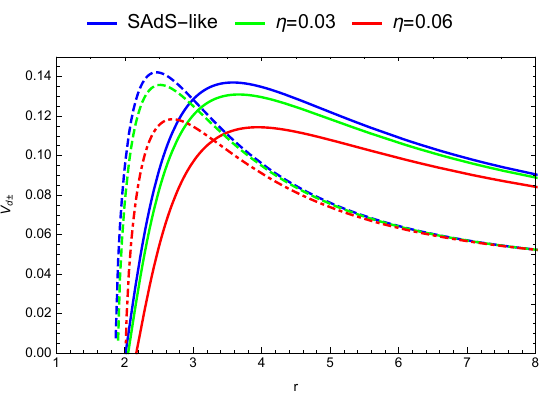}}\label{fig Vdene}}
   \caption{Graphs of the effective potential of the massless Dirac field perturbation: (a) SdS-like black hole ($\Lambda=0.05$) and (b) SAdS-like black hole ($\Lambda=-0.05$) for different values of  global monopole. The physical parameters are chosen as  $M=1$, $l=1$ and $L=0.2$.}
   \label{fig Vde}
\end{figure}

\begin{figure}[h!]
\centering
  \subfloat[\centering  ]{{\includegraphics[width=170pt,height=170pt]{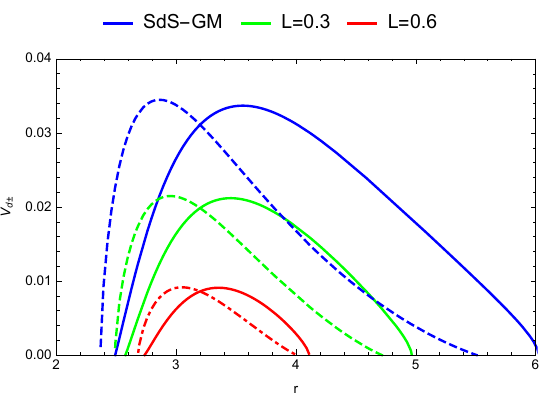}}\label{fig VdLpo}}
  \qquad
   \subfloat[\centering  ]{{\includegraphics[width=170pt,height=170pt]{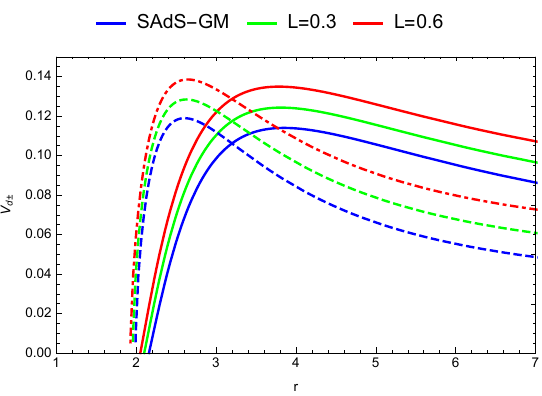}}\label{fig VdLne}}
   \caption{Graphs of the effective potential of the massless Dirac field perturbation: (a) SdS-like black hole ($\Lambda=0.05$) and (b) SAdS-like black hole ($\Lambda=-0.05$) for different values of  Lorentz violation parameter. The physical parameters are chosen as  $M=1$, $l=1$ and $\eta=0.05$.}
   \label{fig VdL}
\end{figure}

The effect of global monopole parameter on the effective potential of massless Dirac field perturbation is illustrated in Fig. \ref{fig Vde}. The solid line represents $V_{d+}$, whereas the dashed line represents $V_{d-}$. Increasing the monopole parameter reduces the maximum of the effective potential in both the cases. This suggests that the monopole parameter  increases the transmission of waves which tends to increase the greybody factor. It is also important to note that increasing the monopole parameter causes the maximum of the effective potential to occur at  a larger radius for both dS and AdS cases. In Fig. \ref{fig VdL}, we highlight the role of Lorentz violation parameter in the effective potential of massless Dirac field perturbation. The Lorentz violation parameter $L$ has opposite effect on the maximum of effective potential of massless Dirac field perturbation for dS and AdS cases. For dS case, increasing $L$ reduces the maximum of the potential and the peak occurs at lower radius while for AdS case, the maximum of the potential increases and the peak occurs at larger radius. This signifies that as the value of $L$ gets higher the greybody factor will be higher or lower according to the sign of $\Lambda$. The above remark will be thoroughly discussed in the subsequent section.


\subsubsection{Massive Dirac field perturbation}
Applying the transformation $Z=\tan ^{-1}\left( \dfrac{\mu_* r}{\lambda}\right)$, Eq. \eqref{eqn dirac radial} reduces to
\begin{align}\label{eqn massive dirac}
\left( \dfrac{d}{dr_*}\pm i\omega\right) P_{\pm} = &\sqrt{1-\kappa \eta^2-\dfrac{2M}{r}-\dfrac{(1+L)\Lambda r^2}{3}}  \exp\left[ \pm i \tan ^{-1}\left( \dfrac{\mu_* r}{\lambda}\right) \right]  \dfrac{(\lambda ^2 +\mu_* ^2 r^2)^\frac{1}{2}}{r} P_{\mp}.
\end{align}
Substituting $P_\pm=U_\pm \exp\left[ \pm\dfrac{i}{2} \tan ^{-1}\left( \dfrac{\mu_* r}{\lambda}\right) \right]$ and changing the variable $r_*$ into $\tilde{r}_{*}$ defined by
\begin{align}
\tilde{ r}_*=r_*+\dfrac{1}{2\omega}\tan ^{-1}\left( \dfrac{\mu_* r}{\lambda}\right) ,
\end{align}
Eq. \eqref{eqn massive dirac} reduces to 
\begin{align}
 \left( \dfrac{d}{d \tilde{r}_*}\pm i\omega\right) U_{\pm} = W U_{\mp},
 \end{align}
 where 
\begin{align}
W=\dfrac{\sqrt{1-\kappa \eta^2-\dfrac{2M}{r}-\dfrac{(1+L)\Lambda r^2}{3}}~ \sqrt{1+L}~(\lambda^2+r^2\mu_*^2)^{\frac{3}{2}}}{r\left[  \sqrt{1+L}~(\lambda^2+r^2\mu_*^2)+\dfrac{\mu_*\lambda}{2\omega}\left(1-\kappa \eta^2-\dfrac{2M}{r}-\dfrac{(1+L)\Lambda r^2}{3}\right)  \right]  }.
\end{align}
Letting $\widetilde{Z}_\pm=U_{-}\pm U_{+}$, we can readily obtain a pair of one dimensional Schr\"{o}ndinger-like wave equations
\begin{eqnarray}
\left( \dfrac{d^2}{d\hat{r}_*^2}+\omega^2\right)\widetilde Z_\pm= V_{dm\pm}\widetilde Z_\pm,
 \end{eqnarray}
  where $ V_{dm \pm}$ are the effective potentials of massive Dirac field perturbation given by
\begin{align}
  V_{dm\pm}(r)&= W^2 \pm \dfrac{dW}{d\hat{r}_*}\nonumber \\
&= \dfrac{\widetilde \Delta^{\frac{1}{2}}\sqrt{1+L} ~ \Xi^{\frac{3}{2}}}{\left[ r^2\sqrt{1+L} ~ \Xi +\dfrac{\widetilde \Delta \mu_* \lambda}{2 \omega}\right] ^2 }+\left[ \widetilde \Delta^{\frac{1}{2}} \sqrt{1+L}~ \Xi ^{\frac{3}{2}} \pm    \left\lbrace 3r\widetilde \Delta \mu_*^{2} +            \Xi\Big(r-          r\kappa \eta ^{2} 
-M-\dfrac{2\Lambda (1+L) r^3}{3}\Big) \right\rbrace \right]  \nonumber\\ &
\mp \dfrac{\widetilde \Delta^{\frac{3}{2}}\sqrt{1+L}~ \Xi^{\frac{5}{2}}}{\left[ r^2\sqrt{1+L}~ \Xi +\dfrac{\widetilde \Delta \mu_* \lambda}{2 \omega}\right] ^3 }\left[ 2r^3 \mu_* ^{2}\sqrt{1+L}        + 2r \sqrt{1+L} ~ \Xi +\dfrac{\mu_* \lambda}{\omega}\left( r-r \kappa \eta ^{2} -M-\dfrac{2\Lambda (1+L) r^3}{3}\right) \right],
\end{align}
where $\widetilde \Delta =r^2\left(1-\kappa \eta^2-\dfrac{2M}{r}-\dfrac{(1+L)\Lambda r^2}{3} \right) $ and $\Xi= (\lambda^2 + r^2\mu_*^{2})$.

\begin{figure}[h!]
\centering
  \subfloat[\centering  ]{{\includegraphics[width=170pt,height=170pt]{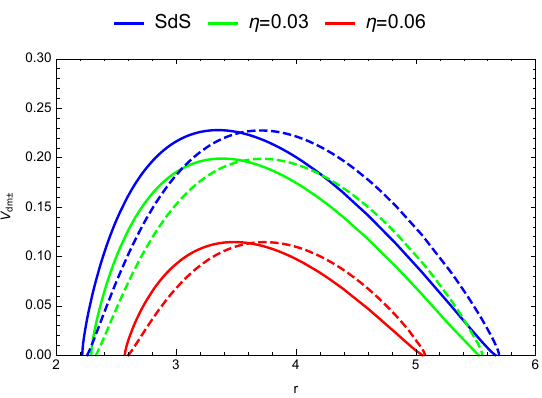}}\label{fig Vdmepo}}
  \qquad
   \subfloat[\centering  ]{{\includegraphics[width=170pt,height=170pt]{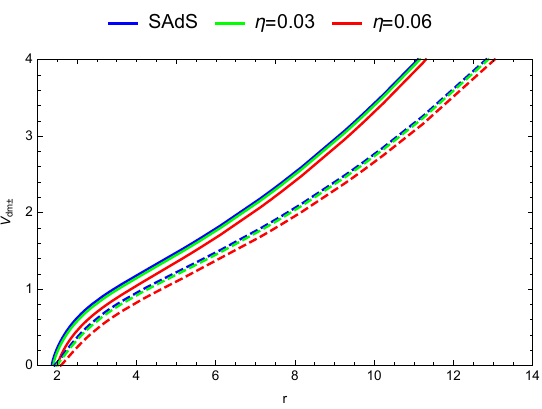}}\label{fig Vdmene}}
   \caption{Graphs of the effective potential of the massive Dirac field perturbation: (a) SdS-like black hole ($\Lambda=0.05$) and (b) SAdS-like black hole ($\Lambda=-0.05$) for different values of  Lorentz violation parameter. The physical parameters are chosen as  $M=1$, $l=1$ and $\eta=0.05$.}
   \label{fig Vdme}
\end{figure}

\begin{figure}[h!]
\centering
  \subfloat[\centering  ]{{\includegraphics[width=170pt,height=170pt]{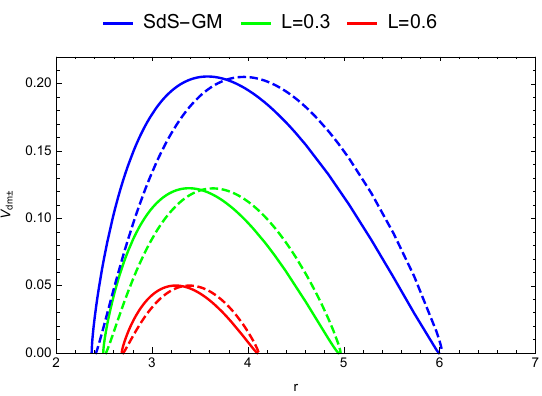}}\label{fig VdmLpo}}
  \qquad
   \subfloat[\centering  ]{{\includegraphics[width=170pt,height=170pt]{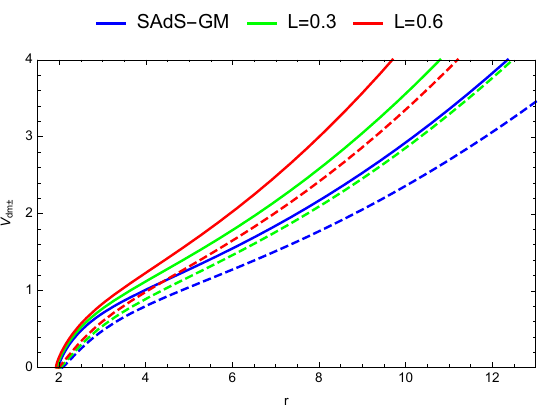}}\label{fig VdmLne}}
   \caption{Graphs of the effective potential of the massive Dirac field perturbation: (a) SdS-like black hole ($\Lambda=0.05$) and (b) SAdS-like black hole ($\Lambda=-0.05$) for different values of  Lorentz violation parameter. The physical parameters are chosen as  $M=1$, $l=1$ and $\eta=0.05$.}
   \label{fig VdmL}
\end{figure}
In Figs. \ref{fig Vdme} and \ref{fig VdmL}, we illustrate the the impact of Lorentz violation parameter and global monopole parameter on the effective potential of massive Dirac field perturbation. In both the dS and AdS cases, increasing the value of global monopole parameter  lowers the peak of the effective potential thereby allowing the transmission of the waves more effectively. However, increasing the Lorentz violation parameter $L$  lowers the peak of the effective potential in the dS case but it rises in the AdS case . For dS case,  increasing the value of $\eta$ shifts the peak of the effective potential towards the right but it shifts towards the left with increasing $L$.

\section{Greybody factor}

During the emission of black hole, some radiation is reflected back into the black hole and others are transmitted out of the black hole. This  modifies the spectrum of Hawking radiation seen at spatial infinity \cite{Hawking1975}. The study of greybody factor is important in understanding the nature of the spectrum of Hawking radiation seen by an observer at a far distance from the event horizon of the black hole. In this section, we will study the rigorous bound of the greybody factor for scalar, electromagnetic and Dirac field perturbations. A rigorous bound of the transmission probability, which is the same as that of the greybody factor, is given by \cite{visser1999,boonserm2008}
\begin{align}\label{eqn grey}
T\geq \sech^2 \left(	 \dfrac{1}{2 \omega} \int_{-\infty}^{+\infty}  \vert V_{eff} \vert ~dr_*	\right).
\end{align}
For dS case, the presence of the event horizon and cosmological horizon provides a well defined setup of studying greybody factors based on wave scattering between two physical horizons. However,  AdS spacetime features a timelike boundary at spatial infinity, which requires careful treatment of boundary conditions. Thus the study of greybody factor using rigorous bound technique in AdS is more subtle, so we restrict our analysis of greybody factor for dS case only.

For dS case, using the definition of $r_*$, the relation for bounds on greybody factors takes the following form
\begin{align}\label{eqn dsgrey}
T\geq \sech^2 \left(	 \dfrac{\sqrt{1+L}}{2 \omega} \int_{r_h}^{r_c} \dfrac{ \vert V_{eff} \vert}{f(r)} ~dr	\right).
\end{align}
\subsection{Scalar perturbation}
Using Eqs. \eqref{eqn f} and \eqref{eqn Vs}, in Eq. \eqref{eqn dsgrey}, we derive the expression for the rigorous bound of the greybody factor of scalar perturbation

\begin{align}\label{eqn sgrey}
T_s \geq \text{sech}^2 &\left[ \dfrac{\sqrt{1+L}}{2\omega}\left\lbrace l(1+l)\left( \dfrac{1}{r_c}-\dfrac{1}{r_e}\right)  -\dfrac{2\Lambda (r_c-r_e)}{3(1+L)} - \dfrac{M}{(1+L)}\left( \dfrac{1}{r_c^{2}}-\dfrac{1}{r_e^{2}}\right)\right\rbrace\right].
\end{align}

\begin{figure}[h!]
\centering
  \subfloat[\centering  ]{{\includegraphics[width=170pt,height=170pt]{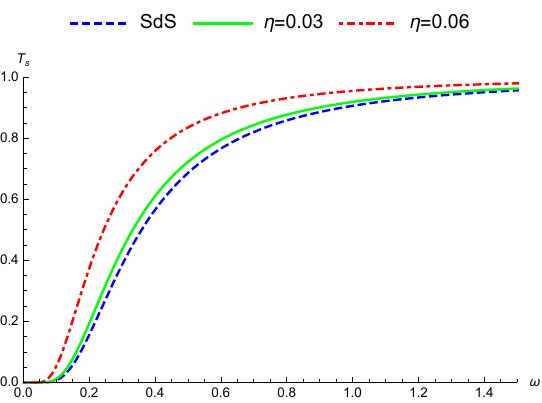}}\label{fig greyse}}
  \qquad
   \subfloat[\centering  ]{{\includegraphics[width=170pt,height=170pt]{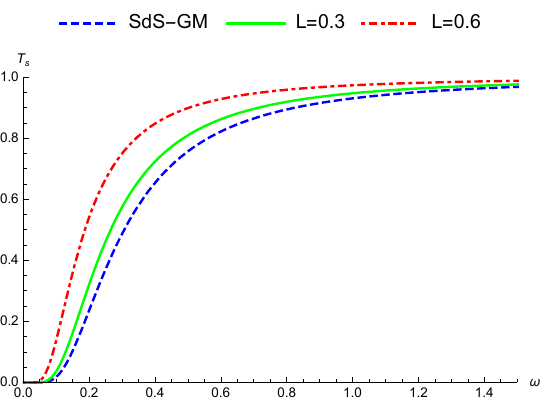}}\label{fig greysL}}
   \caption{Graphs of the lower bound of the greybody factor of scalar field perturbation for different values of  (a) global monopole parameter  and (b) Lorentz violation parameter. The physical parameters are chosen as  $M=1$, $l=1$ and $\Lambda=0.05$.}
   \label{fig greys}
\end{figure}

From Eq. \eqref{eqn sgrey}, one can readily see that the lower bound  of the greybody factor of scalar perturbation depends on the Lorentz violation parameter. Further, the impact of the global monopole parameter can be determined by the indication that the bound depends on the distance between the two horizons. We illustrate the variation of rigorous bound of the greybody factor for electromagnetic perturbation for different values of $\eta$ and $L$ in  Figs. \ref{fig greyse} and \ref{fig greysL} respectively.  One can observe that as the frequency increases,  the lower bound on the greybody factor increases from zero to one. Moreover, increasing the Lorentz violation parameter and global monopole parameter increase the bound of greybody factor. It ultimately results in allowing more waves to reach  a far distance observer.

\subsection{Electromagnetic perturbation}
Using the effective potential of electromagnetic perturbation \eqref{eqn Vm}, we derive the lower bound of the greybody factor 
\begin{align}\label{eqn egrey}
T_s \geq \text{sech}^2 &\left[ \dfrac{ l(1+l)\sqrt{1+L}}{2\omega} \left( \dfrac{1}{r_e}-\dfrac{1}{r_c}\right)  \right].
\end{align}

\begin{figure}[h!]
\centering
  \subfloat[\centering  ]{{\includegraphics[width=170pt,height=170pt]{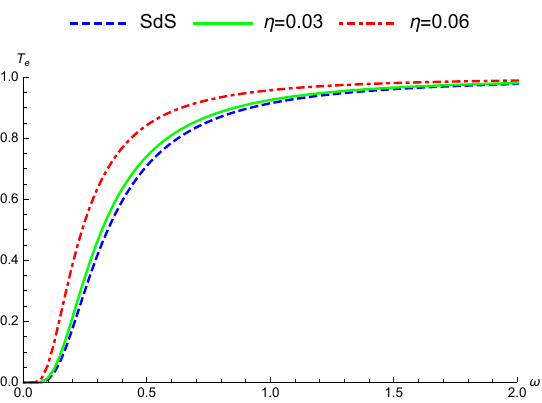}}\label{fig greyee}}
  \qquad
   \subfloat[\centering  ]{{\includegraphics[width=170pt,height=170pt]{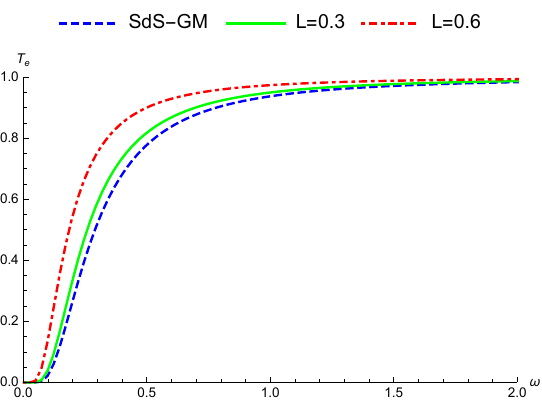}}\label{fig greyeL}}
   \caption{Graphs of the rigorous bound of the greybody factor of electromagnetic field perturbation for different values of  (a) global monopole parameter  and (b) Lorentz violation parameter. The physical parameters are chosen as  $M=1$, $l=1$ and $\Lambda=0.05$.}
   \label{fig Gee}
\end{figure}

Similar to the scalar perturbation, the lower bound of the greybody factor of the electromagnetic perturbation also depends on $L$ and the distance between the horizons. Further, one can understand from Eq. \eqref{eqn egrey} that if the distance between the horizons decrease, the angle in the RHS of \eqref{eqn egrey} also decreases (for a clear point of view, we plot $r_c-r_a$ and $1/r_a-1/r_c$ in Appendix A). Since RHS of \eqref{eqn egrey} is a decreasing function for positive inputs, the lower bound of the greybody factor increases with decreasing the distance between the horizons. Moreover, we already obtained that the distance between the horizons are inversely proportional to $\eta$ and $L$ (see Figs. \ref{fig fpo} and \ref{fig fLpo}). Thus, the rigorous bound of the greybody factor increases with increasing $\eta$ and $L$.  To analyse the effect of $\eta$ and $L$ numerically, we plot the rigorous bound of the greybody factor for different values of $\eta$ and $L$ in Figs. \ref{fig greyee} and \ref{fig greyeL} respectively. The outcomes of the numerical analysis is also consistent with the earlier analysis. 
 
\subsection{Massless Dirac perturbation}
The lower bound of the greybody factor of massless Dirac perturbation is derived as
\begin{align}
T_d \geq \text{sech}^2 \left[ \dfrac{1}{2\omega}\left\lbrace -\sqrt{1+L}\left(l+\dfrac{1}{2}\right)^{2}\left( \dfrac{1}{r_c}-\dfrac{1}{r_e}\right) \pm 
\left(l+\dfrac{1}{2}\right) \left( \dfrac{\sqrt{f(r_c)}}{r_c}- \dfrac{\sqrt{f(r_e)}}{r_e}\right) \right\rbrace \right]. 
\end{align}
Since $f(r)$ vanishes at the horizons, the lower bound of the greybody factor of massless Dirac perturbation reduces to
\begin{align}
T_d \geq \text{sech}^2 \left[ \dfrac{1}{2\omega}\left\lbrace \sqrt{1+L}\left(l+\frac{1}{2}\right)^{2}\left( \dfrac{1}{r_e}-\dfrac{1}{r_c}\right)  \right\rbrace \right]. 
\end{align}

\begin{figure}[h!]
\centering
  \subfloat[\centering  ]{{\includegraphics[width=170pt,height=170pt]{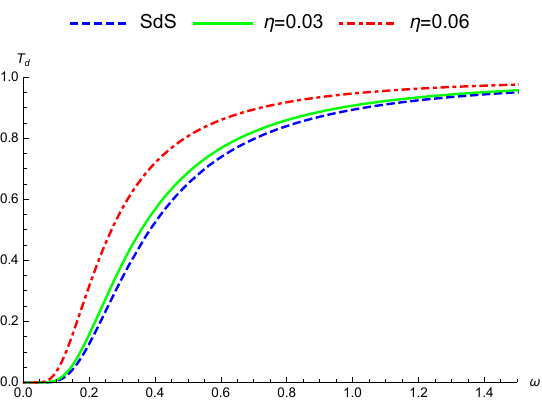}}\label{fig greyde}}
  \qquad
   \subfloat[\centering  ]{{\includegraphics[width=170pt,height=170pt]{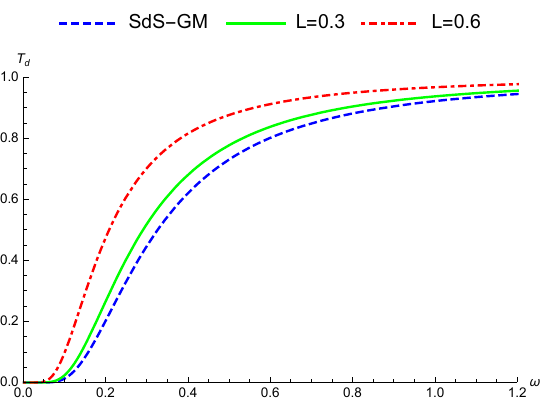}}\label{fig greydL}}
   \caption{Graphs of the lower bound of the greybody factor of massless Dirac field perturbation for different values of  (a) global monopole parameter  and (b) Lorentz violation parameter. The physical parameters are chosen as  $M=1$, $l=1$ and $\Lambda=0.05$.}
   \label{fig Gde}
\end{figure}

The lower bound of the greybody factor of massless Dirac perturbation is similar to that of electromagnetic perturbation. The lower bound of the greybody factor of massless Dirac perturbation for different values of $\eta$ and $L$ are depicted in Figs. \ref{fig greyde} and \ref{fig greydL}. Increasing the values of $\eta$ and $L$ increase the lower bound of the greybody factor thereby allowing more wave to be transmitted and reach a far distance observer.

\subsection{Massive Dirac perturbation}
The lower bound of the greybody factor of massive Dirac perturbation is derived as
\begin{align}\label{eqn greydm}
T_{dm}=\sech^2 \left[ \dfrac{\sqrt{1+L}}{2\omega}    \int_{r_e}^{r_c} 	A \,dr	   + \dfrac{1}{2\omega} \times W\vert_{r_e}^{r_c}\right],		
\end{align}
where \begin{align}
A= \dfrac{\lambda^2 (1+r^2\tilde \mu^2)}{r^2} \left[ 1+ \left(1-\kappa \eta^2-\dfrac{2M}{r}-\dfrac{(1+L)\Lambda r^2}{3}  \right) \times \dfrac{\tilde \mu}{ 2\omega \sqrt{1+L}(1+r^2\tilde \mu^2)} \right]^{-1}, \hspace{0.5cm} \tilde \mu=\dfrac{\mu_*}{\lambda}.
\end{align}
 It is noteworthy to mention that $W$ vanishes at the horizon. Further, it is complicated to evaluate the full integral of Eq. \eqref{eqn greydm}, so we apply  the approximation technique used in \cite{boonserm2021} and calculate the greybody factor bound as
 \begin{align}
 \widetilde T_d \geq \rm{sech}^2 \left[ \dfrac{\sqrt{1+L}}{2\omega}\left\lbrace \mu_*^2(r_c-r_h)-\lambda^2\left( \dfrac{1}{r_c}-\dfrac{1}{r_e}\right)\right\rbrace  \right] .
\end{align}

\begin{figure}[h!]
\centering
  \subfloat[\centering  ]{{\includegraphics[width=170pt,height=170pt]{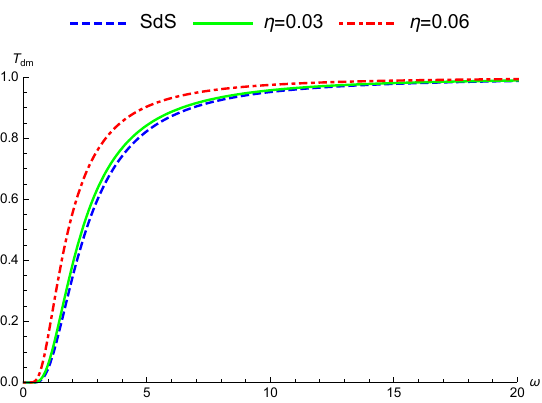}}\label{fig greydme}}
  \qquad
   \subfloat[\centering  ]{{\includegraphics[width=170pt,height=170pt]{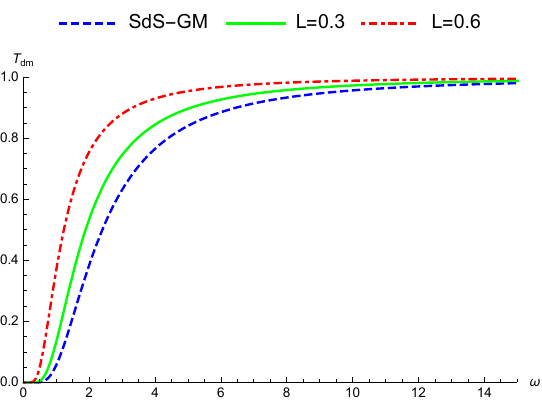}}\label{fig greydmL}}
   \caption{Graphs of the rigorous bound of the greybody factor of massive Dirac field perturbation for different values of  (a) global monopole parameter  and (b) Lorentz violation parameter. The physical parameters are chosen as  $M=1$, $l=1$ and $\Lambda=0.05$.}
   \label{fig Gdm}
\end{figure}

In Fig. \ref{fig Gdm}, we depicted the nature of rigorous bound of the greybody factor of massive Dirac field perturbation. In this case as well, the rigorous bound on the greybody factor rises as the parameters $\eta$ and $L$ increase implying that the resistance to the transmission of outgoing waves is reduced as the above parameters increase. It is evident that for every perturbation,  the bounds asymptotically approach the value 1. 

\section{Quasinormal modes}

In this section, we will calculate the QNM frequencies numerically, using the 6th order WKB approximation along with the improvements figured out using 6th order  Padé Approximation. WKB method was first developed by Schutz and Will \cite{schutz1985} and extended to higher orders in Refs. \cite{S. I C,konoplya2003,matyjasek2017}. The formula for calculating quasinormal frequencies using the 6th order WKB method is given by
\begin{align}
\dfrac{i\left(\omega^2-V(r_0)\right)}{\sqrt{-2 V''(r_0)}}-\Lambda_2-\Lambda_3-\Lambda_4-\Lambda_5-\Lambda_6=n+\dfrac{1}{2},
\end{align}
where $V(r_0)$ and $V''(r_0)$ represent the value of the potential and its second derivative with respect to the tortoise coordinate evaluated at its maxima $r_0$. Here $n$ denotes the overtone number. The expression of $\Lambda_2$, $\Lambda_3$, $\Lambda_4$, $\Lambda_5$, and $\Lambda_6$ can be found in \cite{S. I C,konoplya2003}. WKB method is reliable  and accurate when the overtone number $n$ is small compared to the multipole number $l$. However, when $n>l$ the method becomes less accurate. The precision of the WKB method can be enhanced by incorporating the Pad\'{e} approximation technique introduced in \cite{matyjasek2017,matyjasek2019}. To obtain more accurate and reliable estimates for the QNM frequencies of the black hole in our work, we employ the Pad\'{e} averaged 6th-order WKB approximation. Using the effective potentials of scalar, electromagnetic and Dirac perturbations, we compute the QNM frequencies and display in Table \ref{tab qnms}, \ref{tab qnme} and \ref{tab qnmd} respectively.


\begin{table}[htbp]
    \centering
    \begin{tabular}{c c c c c}      
       Multipole Number & Overtone & \quad \quad    \quad\quad\quad & WKB 6th Order & Padé Approximation  \\      
       \hline
       \hline
    &&  $\eta=0.05$,  $M=1$,   $\Lambda=0.05$\\
       \hline
       & & $L$& &\\
       \hline
       
        $l=1$ & $n=0$ & 0 & 0.166735 - 0.0608741 $i$ &  0.166727 - 0.0608164 $i$   \\
        &  & 0.3 &  0.128337 - 0.0473977 $i$ &  0.128365 - 0.047382 $I$   \\
        &  & 0.6 &  0.0834522 - 0.0300971 $i$ &  0.0827719 - 0.030395 $i$  \\
        
        \hline
        $l=2$ & $n=0$ & 0 &  0.288104 - 0.0581909 $i$ &    0.288104 - 0.0581887 $i$ \\
        &  & 0.3 &  0.227598 - 0.0460515 $i$ &   0.227599 - 0.0460502 $i$ \\
        &  & 0.6 &  0.14898 - 0.0300305 $i$ &   0.148821 - 0.0300655 $i$  \\
        \hline
        $l=2$ & $n=1$ & 0 &  0.28319 - 0.175442 $i$ &   0.28319 - 0.17544 $i$ \\
        &  & 0.3 &  0.224701 - 0.138569 $i$ &  0.224729 - 0.138554 $i$   \\
        &  & 0.6 &  0.149482 - 0.0892945 $i$ &  0.147912 - 0.0902552 $i$   \\
        \hline
        $l=3$ & $n=0$ & 0 &  0.407316 - 0.0575415 $i$ &  0.407316 - 0.0575413 $i$  \\
        &  & 0.3 &  0.323613 - 0.0457497 $i$ &   0.323613 - 0.0457495 $i$ \\
        &  & 0.6 &  0.212272 - 0.0299885 $i$ &  0.212282 - 0.0299877 $i$  \\
         \hline
        $l=3$ & $n=1$ & 0 &  0.403528 - 0.173022 $i$ &  0.403528 - 0.173022 $i$  \\
        &  & 0.3 &  0.32156 - 0.137403 $i$ &  0.321564 - 0.137403 $i$   \\
        &  & 0.6 &  0.211545 - 0.090035 $i$ &  0.21167 - 0.0899829 $i$  \\
         \hline
        $l=3$ & $n=2$ & 0 & 0.396106 - 0.289701 $i$ &  0.396106 - 0.289701 $i$  \\
        &  & 0.3 & 0.317454 - 0.229547 $i$ &   0.317461 - 0.22953 $i$  \\
        &  & 0.6 &  0.209785 - 0.150508 $i$ &  0.210437 - 0.150045 $i$ \\
        \hline
        \hline
    &&  $L=0.2$,  $M=1$,   $\Lambda=0.05$\\
       \hline
       & & $\eta$& &\\
       \hline
        $l=1$ & $n=0$ & 0 &  	0.182786 - 0.0692195	$i$		 &    0.182778 - 0.0691352 $i$			 \\
        & 	 & 0.03 		&  		0.16827 - 0.0631862 $i$		 &   0.168271 - 0.0631013	$i$		 \\
        & 	 & 0.06 		&  		0.121599 - 0.0442161 $i$		&   	0.121644 - 0.0442013 $i$			  \\
         \hline
        $l=2$ & $n=0$ & 0 &  	0.3186195 - 0.0665137 $i$			 &    	0.3186198 - 0.0665106 $i$		 \\
        & 	 & 0.03 		&  		0.2943335 - 0.0607819 $i$	 &    0.2943338 - 0.0607794 $i$ 			 \\
        & 	 & 0.06 		&  		0.2151813 - 0.0428825 $i$		&   	0.2151812 - 0.0428824 $i$			  \\
         \hline
        $l=2$ & $n=1$ & 0 &  		0.312616 - 0.200608 $i$		 &    	0.312614 - 0.200606 $i$		 \\
        & 	 & 0.03 		&  		0.289238 - 0.183226 $i$		 &   	0.289236 - 0.18322 $i$ 		 \\
        & 	 & 0.06 		&  		0.212602 - 0.129005 $i$		&   		0.212602 - 0.129005 $i$
        		  \\
      
         \hline
        $l=3$ & $n=0$ & 0 &  	0.451375 - 0.0658817 $i$			 &    	0.451375 - 0.0658814	$i$	 \\
        & 	 & 0.03 		&  	0.417276 - 0.0602284 $i$			 &   0.417276 - 0.0602281	$i$		 \\
        & 	 & 0.06 		&  		0.3057934 - 0.0425843 $i$		&   	0.3057935 - 0.0425841 $i$			  \\
         \hline
        $l=3$ & $n=1$ & 0 &  	 0.4467949 - 0.198135 $i$			 &   0.4467948 - 0.198134 $i$ 			 \\
        & 	 & 0.03 		&  		0.4134373 - 0.181071	$i$	 &   	0.4134372 - 0.18107 $i$		 \\
        & 	 & 0.06 		&  		0.303949 - 0.12789 $i$		&   	0.303957 - 0.127889	 $i$		  \\
         \hline
        $l=3$ & $n=2$ & 0 &  	0.4378079 - 0.331875 $i$ 			 &    	0.4378075 - 0.331874	$i$	 \\
        & 	 & 0.03 		&  	0.405877 - 0.303093 $i$			 &   0.405876 - 0.303091 $i$			 \\
        & 	 & 0.06 		&  	0.300251 - 0.213639 $i$			&   	0.300281 - 0.213602	$i$		  \\
        
        \hline
    \end{tabular}
    \caption{QNM frequencies for scalar perturbation for the SdS-like black hole with global monopole calculated using 6th order WKB and 6th order Padé Approximation for different modes and for different values of the Lorentz violation parameter and monopole parameter.}
    \label{tab qnms}
\end{table}


%
\begin{table}[htbp]
    \centering
    \begin{tabular}{c c c c c}
       
        Multipole Number & Overtone & \quad \quad    \quad\quad\quad & WKB 6th Order & Padé Approximation  \\      
       \hline
       \hline
    &&  $\eta=0.05$,  $M=1$,   $\Lambda=0.05$\\
       \hline
       & & $L$& &\\
       \hline
        $l=1$ & $n=0$ & 0 &  	0.154693 - 0.0557613 $i$			 &    0.154695 - 0.0557307	$i$		 \\
        & 	 & 0.3 		&  	0.124014 - 0.0448676	$i$		 &   0.124022 - 0.0448217 $i$		 \\
        & 	 & 0.6 		&  	0.0799587 - 0.030461	$i$		&   	0.0819019 - 0.0297455	$i$		  \\
         \hline
        $l=2$ & $n=0$ & 0 &  	0.281284 - 0.0565098	$i$		 &    0.281284 - 0.0565086	$i$		 \\
        & 	 & 0.3 		&  		0.2249958 - 0.0452443 $i$		 &   0.2249957 - 0.0452439 $i$			 \\
        & 	 & 0.6 		&  			0.148313 - 0.0298338	$i$    &   	0.148241 - 0.0298501	$i$		  \\
         \hline
        $l=2$ & $n=1$ & 0 &  	0.275597 - 0.170178 $i$			 &    0.275596 - 0.170178 $i$			 \\
        & 	 & 0.3 		&  		0.2221306 - 0.135928	$i$	 &   0.2221308 - 0.135932 $i$		 \\
        & 	 & 0.6 		&  		0.148154 - 0.0891358	$i$	&   	0.14744 - 0.0895641 $i$			  \\
        
         \hline
        $l=3$ & $n=0$ & 0 &  		0.402504 - 0.0567037	$i$	 &    0.402504 - 0.0567036	$i$		 \\
        & 	 & 0.3 		&  	0.321751 - 0.0453484	$i$		 &   	0.321751 - 0.0453483	$i$	 \\
        & 	 & 0.6 		&  	0.211872 - 0.029877 $i$		&   		0.211859 - 0.0298794	$i$	  \\
         \hline
        $l=3$ & $n=1$ & 0 &  		0.398492 - 0.1704388	$i$	 &    	0.398492 - 0.1704387	$i$	 \\
        & 	 & 0.3 		&  	0.319716 - 0.136148 $i$		 &   	0.319716 - 0.136149 $i$		 \\
        & 	 & 0.6 		&  		0.211429 - 0.0895877	$i$	&   	0.211285 - 0.089648	 $i$		  \\
         \hline
        $l=3$ & $n=2$ & 0 &  	0.390427 - 0.285232	$i$		 &   0.390438 - 0.285227 $i$			 \\
        & 	 & 0.3 		&  	0.315603 - 0.227278	$i$		 &   	0.315603 - 0.227279	$i$	 \\
        & 	 & 0.6 		&  	0.210875 - 0.148929 $i$			&   		0.210125 - 0.149455 $i$		  \\
         \hline
        \hline
    &&  $L=0.2$,  $M=1$,   $\Lambda=0.05$\\
       \hline
       & & $\eta$& &\\ \hline
        $l=1$ & $n=0$ & 0 &  		0.170869 - 0.0638577	$i$	 &    	0.170874 - 0.0638116	$i$	 \\
        & 	 & 0.03 		&  		0.15851 - 0.0585237 $i$	 &   0.158513 - 0.0584857 $i$			 \\
        & 	 & 0.06 		&  		0.117419 - 0.0417628	$i$	&   	0.117419 - 0.0417425 $i$			  \\
         \hline
        $l=2$ & $n=0$ & 0 &  	0.311815 - 0.0647804	$i$		 &   0.311815 - 0.0647786 $i$			 \\
        & 	 & 0.03 		&  	0.288715 - 0.0592863	$i$		 &   0.288715 - 0.0592853	$i$		 \\
        & 	 & 0.06 		&  	0.21265 - 0.0421061 $i$			&   		0.21265 - 0.0421056	$i$	  \\
         \hline
        $l=2$ & $n=1$ & 0 &  		0.304972 - 0.195159 $i$		 &    	0.304973 - 0.195158 $i$	 \\
        & 	 & 0.03 		&  	0.283058 - 0.178463 $i$			 &   	0.283059 - 0.178464 $i$		 \\
        & 	 & 0.06 		&  	0.210086 - 0.126487 $i$			&   		0.210086 - 0.126488	$i$	  \\
        
         \hline
        $l=3$ & $n=0$ & 0 &  	0.446564 - 0.0650204 $i$			 &    0.446564 - 0.0650202 $i$			 \\
        & 	 & 0.03 		&  	0.413295 - 0.0594857 $i$			 &   	0.413295 - 0.0594856 $i$ 		 \\
        & 	 & 0.06 		&  	0.303987 - 0.0421981 $i$			&   	0.303987 - 0.042198 $i$ 			  \\
         \hline
        $l=3$ & $n=1$ & 0 &  	0.441738 - 0.195474 $i$			 &    0.441738 - 0.195474 $i$			 \\
        & 	 & 0.03 		&  	0.4093 - 0.178765 $i$			 &   0.4093 - 0.178765 $i$			 \\
        & 	 & 0.06 		&  	0.302167 - 0.126683 $i$			&   		0.302167 - 0.126683 $i$		  \\
         \hline
        $l=3$ & $n=2$ & 0 &  	0.432041 - 0.327268 $i$			 &    		0.432057 - 0.327259 $i$	 \\
        & 	 & 0.03 		&  	0.401254 - 0.299042	$i$		 &   0.40126 - 0.299042	 $i$		 \\
        & 	 & 0.06 		&  	0.298487 - 0.211453	$i$		&   		0.298487 - 0.211453	$i$	  \\
         \hline
       
    \end{tabular}
    \caption{QNM frequencies for electromagnetic perturbation for the SdS-like black hole with global monopole calculated using 6th order WKB and 6th order Padé Approximation for different modes and for different values of the Lorentz violation parameter and monopole parameter.}
    \label{tab qnme}
\end{table}

\begin{table}[htbp]
    \centering
    \begin{tabular}{c c c c c}
       
        Multipole Number & Overtone & \quad \quad    \quad\quad\quad & WKB 6th Order & Padé Approximation  \\      
       \hline
       \hline
    &&  $\eta=0.05$,  $M=1$,   $\Lambda=0.05$\\
       \hline
       & & $L$& &\\
       \hline
        $l=1$ & $n=0$ & 0 &  	0.174652 - 0.0567557	$i$		&    	0.17465 - 0.0567437	$i$	 \\
        & 	 & 0.3 		&  			0.139148 - 0.0451919 $i$	 &   0.138581 - 0.0453736 $i$			 \\
        & 	 & 0.6 		&  		0.125657 - 0.0228762	$i$	&   		0.0908757 - 0.0299374 $i$		  \\
         \hline
        $l=2$ & $n=0$ & 0 &  		0.292869 - 0.0568447 $i$		 &    0.292868 - 0.0568451 $i$			 \\
        & 	 & 0.3 		&  		0.233379 - 0.0453983	$i$	 &   	0.233328 - 0.0454091 $i$ 		 \\
        & 	 & 0.6 		&  		0.151933 - 0.0303426	$i$	&   		0.153298 - 0.0298997 $i$		  \\
         \hline
        $l=2$ & $n=1$ & 0 &  		0.287301 - 0.171186 $i$		 &    	0.287294 - 0.171195	$i$	 \\
        & 	 & 0.3 		&  	0.231081 - 0.136086 $i$			 &   	0.230861 - 0.136485	$i$	 \\
        & 	 & 0.6 		&  	0.140511 - 0.0994076	$i$		&   		0.152387 - 0.0898092	$i$	  \\
         \hline
        $l=3$ & $n=0$ & 0 &  		0.410708 - 0.0568714	$i$  &   	0.327625 - 0.0454266 $i$		 \\
         & 	 & 0.3 		&  			0.327608 - 0.0454285	$i$ &   	0.327625 - 0.0454266 $i$		 \\
        & 	 & 0.6 		&  		0.218379 - 0.029537 $i$	&   	0.21543 - 0.0299024 $i$			  \\
         \hline
        $l=3$ & $n=1$ & 0 &  		0.406732 - 0.170944	$i$	 &    	0.406732 - 0.170945 $i$		 \\
        & 	 & 0.3 		&  	0.325417 - 0.136461 $i$			 &   0.325604 - 0.136383	$i$		 \\
        & 	 & 0.6 		&  		0.247312 - 0.078325	$i$	&   		0.214863 - 0.0897147	$i$	  \\
         \hline
        $l=3$ & $n=2$ & 0 &  		0.398756 - 0.286073	$i$	 &    0.39876 - 0.286074 $i$			 \\
        & 	 & 0.3 		&  	 0.320559 - 0.228353	$i$		 &   	0.321528 - 0.227676 $i$	 \\
        & 	 & 0.6 		&  		0.376022 - 0.0863962	$i$	&   		0.213723 - 0.149558 $i$		  \\      
        \hline
        \hline
    &&  $L=0.2$,  $M=1$,   $\Lambda=0.05$\\
       \hline
       & & $\eta$& &\\
       \hline
        $l=1$ & $n=0$ & 0 &  		0.192324 - 0.0650849 $i$ 		 &  0.192317 - 0.0650785 $i$    			 \\
        & 	 & 0.03 		&  		0.17813 - 0.0595326 $i$		 &   0.178126 - 0.0595232 $i$			 \\
        & 	 & 0.06 		&  	0.130443 - 0.0425033	$i$		&   		0.131332 - 0.0422138	$i$	  \\
         \hline
        $l=2$ & $n=0$ & 0 &  		0.32414 - 0.0651743 $i$		 &    	0.324138 - 0.0651765 $i$		 \\
        & 	 & 0.03 		&  	0.299988 - 0.0596109 $i$			 &   0.299987 - 0.0596111	$i$	 \\
        & 	 & 0.06 		&  	0.220719 - 0.04224 $i$			&   		0.220657 - 0.0422537	$i$	  \\
         \hline
        $l=2$ & $n=1$ & 0 &  			0.3175 - 0.196308 $i$	 &    0.317449 - 0.196347	$i$		 \\
        & 	 & 0.03 		&  		0.294452 - 0.179434	$i$	 &   	0.294447 - 0.179443 $i$		 \\
        & 	 & 0.06 		&  		0.218732 - 0.126573	$i$	&   	0.218124 - 0.12693	$i$		  \\
         \hline
        $l=3$ & $n=0$ & 0 &  		0.455254 - 0.0652122	$i$	 &   0.455254 - 0.0652119 $i$ 			 \\
        & 	 & 0.03 		&  		0.42125 - 0.0596431 $i$		 &   0.421249 - 0.0596435	$i$		 \\
        & 	 & 0.06 		&  	0.309644 - 0.0422681	$i$		&   	0.309639 - 0.0422692 $i$			  \\
         \hline
        $l=3$ & $n=1$ & 0 &  	0.450474 - 0.1960514	$i$		 &    0.450475 - 0.1960511	$i$	 \\
        & 	 & 0.03 		&  		0.417304 - 0.179233	$i$	 &   		0.417292 - 0.179241 $i$	 \\
        & 	 & 0.06 		&  		0.307895 - 0.12687 $i$		&   	0.307867 - 0.126895 $i$			  \\
         \hline
        $l=3$ & $n=2$ & 0 &  	0.440867 - 0.328238 $i$			 &   0.440892 - 0.328225 $i$ 			 \\
        & 	 & 0.03 		&  		0.409391 - 0.29978 $i$		 &  0.409377 - 0.299805 	$i$		 \\
        & 	 & 0.06 		&  		0.304512 - 0.211582	$i$	&   	0.304183 - 0.211814	 $i$		  \\
       \hline 
    \end{tabular}
    \caption{QNM frequencies for massless Dirac perturbation for the SdS-like black hole with global monopole calculated using 6th order WKB and 6th order Padé Approximation for different modes and for different values of the Lorentz violation parameter and monopole parameter.}
    \label{tab qnmd}
\end{table}

In all the three types of perturbations, for a fixed multipole and overtone number, both the real part  and  the magnitude of the imaginary part of the QNM frequency decrease with increasing the Lorentz violation parameter $L$. This implies that the oscillation frequency  and the damping rate decrease  with increasing $L$. The same behaviours are obtained when $\eta$ increases by fixing other parameters. 
Further, for fixed values of $l$, $\eta$ and $L$, the real part of the quasinormal frequency decreases while the magnitude of imaginary part increases  as $n$ increase. Thus with increasing $n$, the oscillation frequency decreases but the damping rate increases. Further with increasing $l$ and fixing other parameters, the  oscillation frequency increases in all the perturbations. However, for varying $l$, the damping rate has different nature in different types of perturbation. With increasing $l$, the damping rate for scalar perturbation decreases but for electromagnetic perturbation it increases.  For  massless Dirac perturbation, the damping rate decreases in all the modes except for $n=0$.

%
%
%

\section{Shadow radius of black hole}

In this section, we study the role of Lorentz violation parameter $L$ and the monopole parameter $\eta$ on the shadow radius of  spherically symmetric black hole with global monopole in bumblebee gravity. For the  metric \eqref{eqn line}, the Lagrangian is expressed as
\begin{align}\label{eqn lagran}
2 \mathscr{L}= f(r) \dot{t}^2 -\dfrac{(1+L) \dot{r}^2}{f(r)}- r^2 \dot{\theta}^2 - r^2 \sin^2\theta \dot{\varphi}^2,
\end{align}
where the overdot symbol represents the derivative with respect to the proper time $\tau$. Without loss of generality, we restrict our analysis only to the equatorial plane i.e. $\theta=\frac{\pi}{2}$. The generalized momenta are derived from the Lagrangian \eqref{eqn lagran} as

\begin{align}\label{eqn momenta}
p_t=f(r) \dot{t}=E, \hspace{1cm} p_r=\dfrac{1+L}{f(r)} \dot{r}, \hspace{1cm} p_\phi=-r^2=-\Lb,
\end{align}
where $E$ and $\Lb$ are the energy and angular momentum of the particle. Using Eq. \eqref{eqn momenta}, Eq. \eqref{eqn lagran} for null geodesic can be written as 
\begin{align}
\dot{r}^2+V=0,
\end{align}
where 
\begin{align}\label{eqn V}
V=\dfrac{1}{1+L} \left(\dfrac{f(r) \Lb^2}{r^2}-E^2 \right).
\end{align}
 The null-like geodesics of the equatorial circular motion should satisfy the conditions
\begin{align}\label{eqn Vrp}
V(r)\vert_{r=r_p}=0 \hspace{0.5cm} \text{and} \hspace{0.5cm}  V'(r)\vert_{r=r_p}=0,
\end{align}
where $r_p$ is the radius of the photon sphere. The critical impact parameter for the photon sphere is
\begin{align}
b_c=\dfrac{\Lb}{E}=\dfrac{r_p}{\sqrt{f(r_p)}}.
\end{align}
 From Eq. \eqref{eqn Vrp},  one can obtain the radius of the photon sphere as
\begin{align}
r_p=\dfrac{3 M}{ (1-\kappa \eta^2)}.
\end{align}

\begin{figure}
\centering
\includegraphics[scale=0.6]{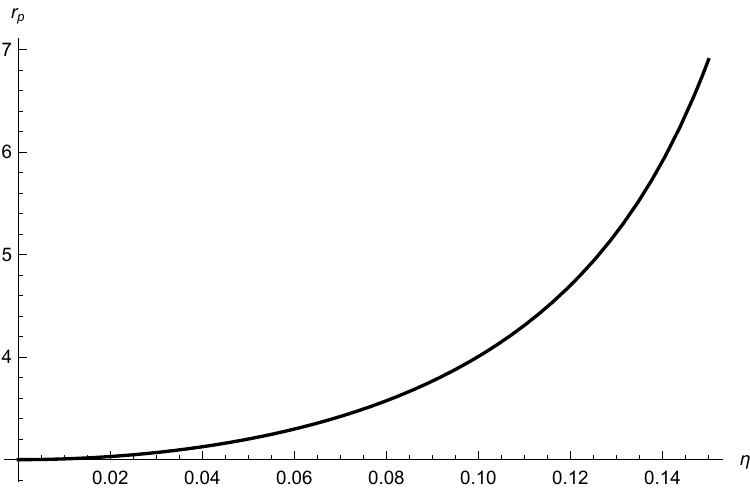}
\caption{Plot of photon radius for different values of $\eta$ and fixed $M=1$.}
\label{fig rp}
\end{figure}

From the expression of $r_p$, one can observe that the monopole parameter has a significant impact on the radius of the photon sphere. However, the Lorentz violation does not affect it. The photon radius is illustrated in Fig. \ref{fig rp}.
Further, the radius of the black hole shadow observed from a distant static observer at the position $r_0$ \cite{konoplya2020} is found to be
\begin{align}
\mathcal{R}_s=r_p \sqrt{\dfrac{f(r_0)}{f(r_p)}}.
\end{align}

Assuming the observer is sufficiently far away from the black hole, we can consider $f(r_0) \approx 1$. Thus the shadow radius reduces to
\begin{align}
\mathcal{R}_s= \dfrac{r_p}{\sqrt{f(r_p)}}.
\end{align}
Further, the shadow radius can also be represented by the celestial coordinate $(X,Y)$ as
\begin{align}
\mathcal{R}_s=\sqrt{X^2+Y^2},
\end{align}
where $ X= \lim\limits_{r_0 \to \infty} \left(-r_{0}^2  \sin\theta_0 \frac{d\phi}{dr}\vert_{r_0,\theta_0} \right)  \text{and}~ Y= \lim\limits_{r_0 \to \infty} \left(r_{0}^2  \frac{d\phi}{dr}\vert_{r_0,\theta_0} \right)$. Here $(r_0,\theta_0)$ denotes the positions of the observer at spatial infinity.
For equatorial plane, the shadow radius is equivalent to the critical impact parameter of the photon sphere. Thus the shadow radius is expressed as
\begin{align}
\mathcal{R}_s= \dfrac{r_p}{\sqrt{1-\kappa\eta^2 -\dfrac{2M}{r_p}-\dfrac{ (1+L)\Lambda r_{p}^2}{3}}}.
\end{align}

\begin{figure}[h!]
\centering
  \subfloat[\centering  ]{{\includegraphics[width=200pt,height=170pt]{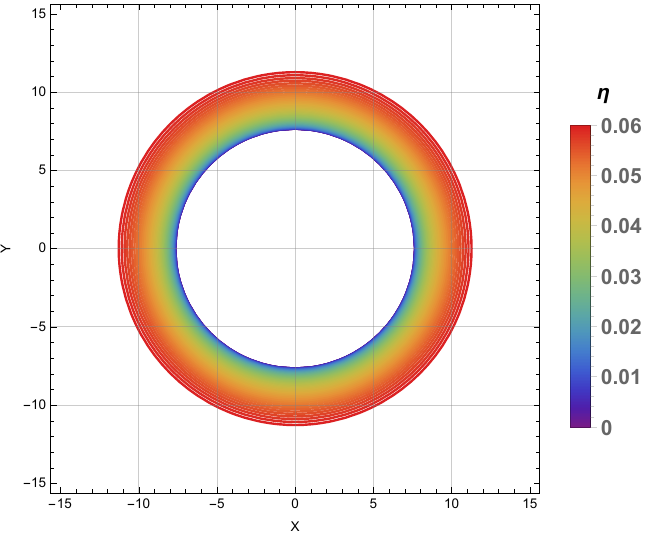}}\label{fig shadowep}}
  \qquad
   \subfloat[\centering  ]{{\includegraphics[width=200pt,height=170pt]{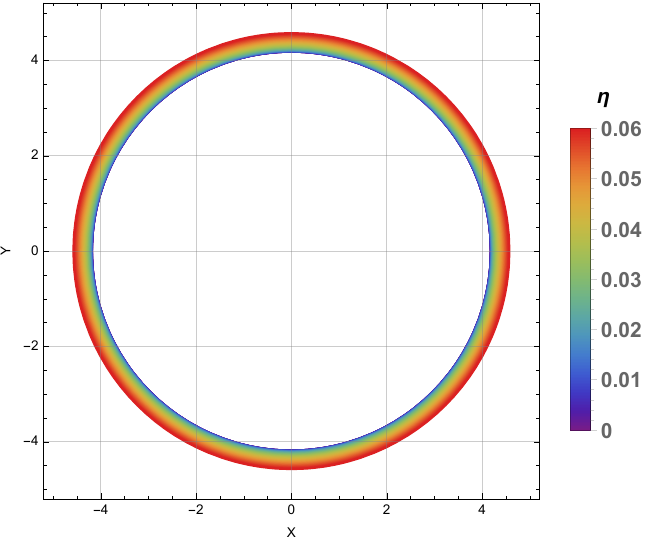}}\label{fig shadowen}}
   \caption{Plot of the shadow radius of (a) SAdS-like black hole with global monopole (b) SdS-like black hole with global monopole  for different values of  global monopole parameter with fixed $L=0.2$, $M=1$ and $\Lambda=0.05$.}
   \label{fig shadowe}
\end{figure}

\begin{figure}[h!]
\centering
  \subfloat[\centering  ]{{\includegraphics[width=200pt,height=170pt]{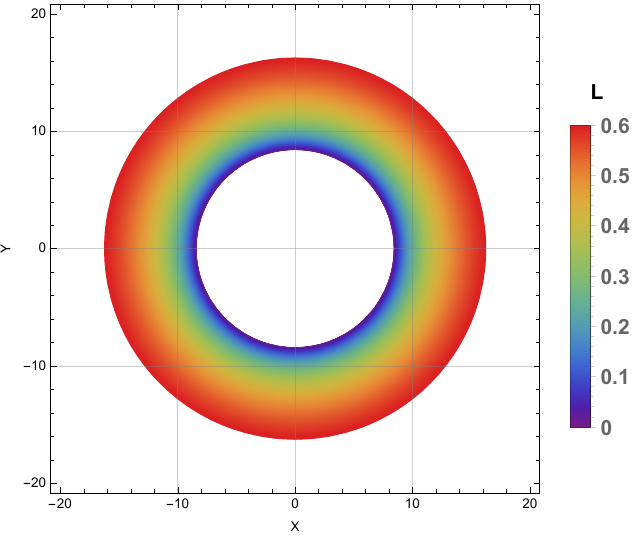}}\label{fig shadowLp}}
  \qquad
   \subfloat[\centering  ]{{\includegraphics[width=200pt,height=170pt]{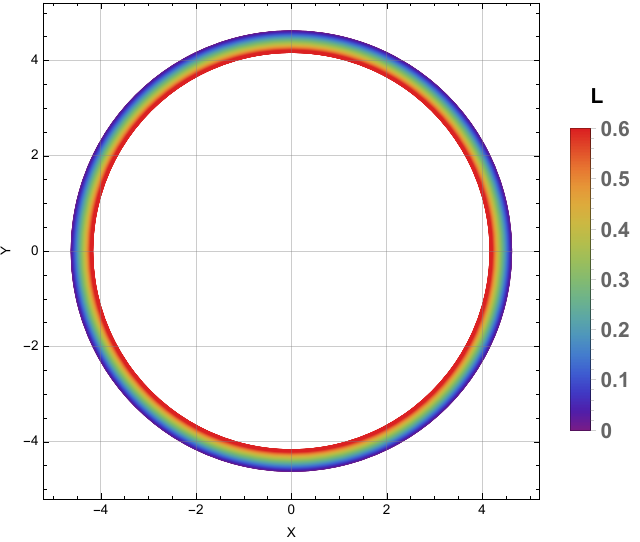}}\label{fig shadowLn}}
   \caption{Plot of the shadow radius of (a) SdS-like black hole with global monopole (b) SAdS-like black hole with global monopole  for different values of  Lorentz violation parameter $L$ with fixed $\eta=0.05$, $M=1$ and $\Lambda=0.05$.}
   \label{fig shadowL}
\end{figure}
To analyse the qualitative nature of variation of  black hole shadow radius, we plot  $\mathcal{R}_s$  for different values of $\eta$ and $L$ in Figs. \ref{fig shadowe} and \ref{fig shadowL} respectively. 
 It is observed that shadow radius increases with increasing the parameters  $\eta$ for both the dS and AdS cases. However with increasing the Lorentz violation parameter, the shadow radius may increase or decrease according to dS or AdS case. 
The effects of Lorentz violation parameter and monopole parameter on the photon radius and shadow radius of dS and AdS spacetimes are shown in tables \ref{tab shadow} and \ref{tab shadow1} respectively. For fixed $L$, both $r_p$ and $R_s$ increase with  increasing $\eta$. However for fixed $\eta$, $r_p$ remains constant and $R_s$ increases with increasing $L$.
\begin{table}[h]
 \centering
    \begin{tabular}{c| c c c c| c c c c}
    $\eta$   &  0 &   0.03 &  0.06 & 0.08 & 0.05 &  0.05 & 0.05 &0.05  \\  
\hline
$L$ &   0.2 &   0.2 &  0.2 & 0.2 &   0 &  0.3 &  0.6 & 0.8\\

$r_p$    &    3 &   3.06943 &   3.29844  & 3.57504 &   3.20113  &  3.20113  & 3.20113 & 3.20113\\  
$R_s$  &   7.66131  &  8.28168 &   11.2751   & 23.0298&   8.50686   &   10.6488   &   16.1828 & 45.3991
 \end{tabular}
\caption{Various values of photon radius and shadow radius for different values of $\eta$ and $L$ with fixed M=1 and $\Lambda=0.05$.}
    \label{tab shadow}
\end{table}
\begin{table}[h]
 \centering
    \begin{tabular}{c| c c c c| c c c c}
    $\eta$   &  0 &   0.03 &  0.06 & 0.08 & 0.05 &  0.05 & 0.05 &0.05  \\  
\hline
$L$ &   0.2 &   0.2 &  0.2 & 0.2 &   0 &  0.3 &  0.6 & 0.8\\

$r_p$    &    3 &   3.06943 &   3.29844  & 3.57504 &   3.20113  &  3.20113  & 3.20113 & 3.20113\\  
$R_s$  &   4.18718  &  4.28038 &   4.57073   & 4.88617  &   4.60522   &   4.3789   &   4.18297 & 4.06608
 \end{tabular}
\caption{Various values of photon radius and shadow radius for different values of $\eta$ and $L$ with fixed M=1 and $\Lambda=-0.05$.}
    \label{tab shadow1}
\end{table}
It is worth mentioning that measurements of the shadow size around the black hole may help to estimate the black hole parameters and also probe the
geometry of the background metric.

\section{ High-energy absorption cross-section}
The study of absorption cross-section of a black hole is one of the  interesting aspects in investigating the field perturbation around the black hole. The absorption cross-section of a black hole is the measure of how much    the incoming particles is absorbed by the black hole rather than dispersed. In this section, we calculate the high-energy absorption cross-section using the sinc approximation method. Sanchez \cite{sanchez1978} proposed that in the high-frequency regime, the total absorption cross-section  oscillated around the constant geometric-optics value for the black hole. For low-energy regime, the absorption cross-section is equal to the area of black hole at the horizon \cite{das1997}. 
 Further, Decanini et al. \cite{decanini2011} used the Regge pole techniques to prove the oscillatory part of the absorption cross-section in the high-frequency regime is related to a sinc(x) function including the photon sphere. In the sinc approximation, the oscillatory part of the absorption cross-section
is given by
\begin{align}
\sigma_{osc}= -8 \pi \sigma_{geo}\, n_c \exp[-\pi n_c]~ \sinc\left[\dfrac{2\pi \omega}{\Omega_c}\right],
\end{align}
where $\sinc z=\sin z/z$ is the sine cardinal, $\sigma_{geo}=\pi b_{c}^2$ is the geometrical absorption cross-section, $\Omega_p=\sqrt{\frac{f_p}{r_{p}^2}}$ is the orbital angular velocity, $r_p$ is the radius of the unstable null circular orbit and $n_c$ is a factor related to the Lyapunov exponent $\lambda_p$ at the unstable circular orbits radius \cite{decanini2010}
\begin{align}
n_c=\dfrac{\lambda_p}{\Omega_p}.
\end{align}
The Lyapunov exponent is given by the relation 
\begin{align}
\lambda_p=\sqrt{-\dfrac{V''}{2\dot{t}^2}}.
\end{align}
Using Eq. \eqref{eqn V}, we derived the expression as
\begin{align}
\lambda_p=& \dfrac{f(r_p)}{2(1+L) r_{p}^2} \left[2 f(r_p)-r_{p}^2 f''(r_p)\right] \nonumber\\
=& \dfrac{(1-\kappa \eta^2) \left(  1-3 \kappa \eta^2+3\kappa^2 \eta^4    - \kappa^3 \eta^6-9(1+L) M^2 \Lambda \right)}{27 (1 + L) M^2}.
\end{align}

Then the total absorption cross-section in the limit for high
frequencies using $\sinc$ approximation  is given by 
\begin{align}
\sigma_{abs} \approx \sigma_{osc}+\sigma_{geo}.
\end{align}

\begin{figure}[h!]
\centering
  \subfloat[\centering  ]{{\includegraphics[width=170pt,height=170pt]{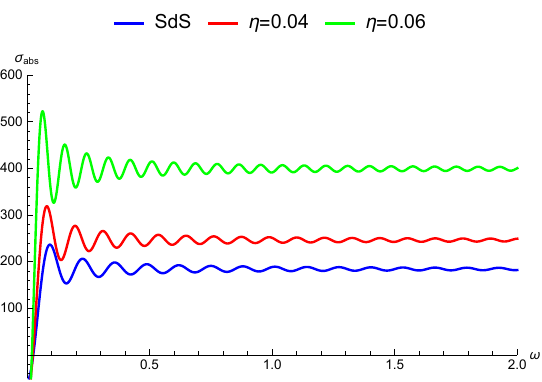}}\label{fig sigmaep}}
  \qquad
   \subfloat[\centering  ]{{\includegraphics[width=170pt,height=170pt]{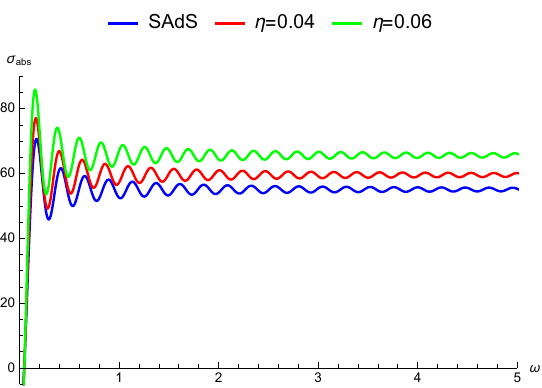}}\label{fig sigmaen}}
   \caption{Graphs of the total absorption cross-section for different values of $\eta$.}
   \label{fig abse}
\end{figure}

\begin{figure}[h!]
\centering
  \subfloat[\centering  ]{{\includegraphics[width=170pt,height=170pt]{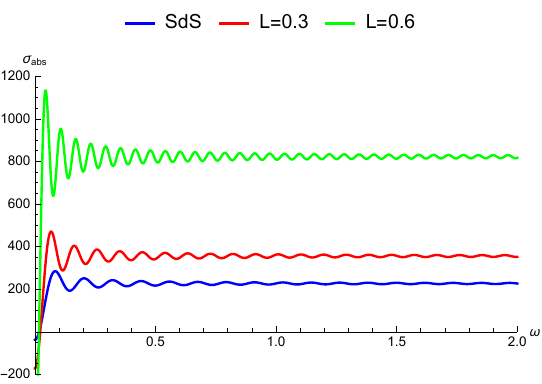}}\label{fig sigmaLp}}
  \qquad
   \subfloat[\centering  ]{{\includegraphics[width=170pt,height=170pt]{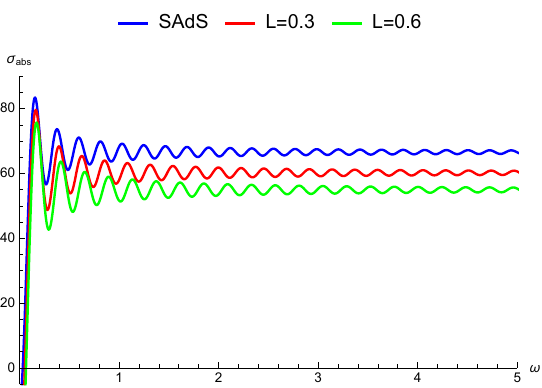}}\label{fig sigmaLn}}
   \caption{Graphs of the total absorption cross-section for different values of $L$ with fixed $M=1$.}
   \label{fig absL}
\end{figure}
We plot the total absorption cross-section for various values of $\eta$ and $L$ in Figs. \ref{fig abse} and \ref{fig absL} respectively. In all the graphs, we observe the three stages of absorption cross-section: (i) the fast growing phase at low frequency, (ii) the intermediate oscillating phase which is a characteristic of interference due to potential barriers and (iii) finally at high frequency the cross section starts to level off. With increasing the monopole parameter, the absorption cross-section becomes higher in both the dS and AdS cases. The absorption cross-section will increase with increasing $L$ in dS case but it will be an opposite effect in AdS black hole.

\section{Conclusion}
In this work, we  investigated scalar,  electromagnetic  and Dirac perturbations of  SdS/AdS-like black hole with a global monopole in bumblebee gravity. We analysed the effective potentials of the above three perturbations and studied the effects of $\eta$ and $L$ in the effective potential, greybody factors, QNMs, shadow radius and absorption cross-section of the black hole. In the region between the event horizon and cosmological horizon, the nature of effective potentials are same for all the perturbations.   For both dS and AdS cases, increasing the   $\eta$  leads to a lower potential barrier. However the $L$ has an opposite effects on the effective potential in dS and AdS backgrounds. Increasing $L$ lowers the potential barrier in the dS case but  it has an opposite  effect in AdS case. Our analysis of greybody factor is limited to the dS only.  The greybody factor is analysed using the rigorous bound technique. In all the perturbations, we found that the lower bounds of the greybody factor increases with increasing $L$  and $\eta$. Thus, increasing the $\eta$  and $L$ facilitate the ability of waves to propagate through the black hole potential and reach asymptotic observers. This behaviour is consistent with the fact that higher values of $L$ and $\eta$ correspond to the lower potential making it easier for waves to transmit through the barrier.

We also derived numerically, the QNMs of dS case by applying the
6th order WKB method and Padé method for varying $\eta$ and $L$. Increasing the parameters $L$ and $\eta$ reduce both the oscillation frequency and the  decay rate. For any value of $L$ and $\eta$, we found no modes with positive damping in all the perturbations. This indicates the stability of SdS black hole perturbation in the presence of  global monopole and Lorentz violation. The photon sphere radius and  black hole shadow radius are also discussed.  Our analysis reveals that with increasing $\eta$, the photon sphere expands  and an enlarged black hole shadow is found for both dS and AdS spacetimes. It is noted that increasing the value of $L$ prevents the rise of shadow
radius in AdS black hole but it has an opposite effect for dS black hole. We also examine the absorption cross-section of the black hole and  find that it
rises with increasing values of $\eta$ for both dS and AdS black holes. However, the role of $L$ again differs between the two spacetimes. Higher $L$ reduces the absorption cross-section in AdS spacetime but it increases the cross-section in a dS background.

Our analysis will enhance the understanding of how the global monopole parameter and Lorentz violation theory  influence black hole observables such as QNM frequencies, shadow profile and thermal radiation properties of black holes. This will contribute to a deeper theoretical understanding of modified gravity scenarios and provide insight into how topological defects and Lorentz-violating effect the future gravitational wave signals and black hole imaging data. The present work excludes the study of gravitational lensing and time-domain analysis of QNMs, which are crucial for a more complete understanding of black hole dynamics and observational signatures. We intend to investigate these in  near future.

\section*{Appendix A}

\begin{figure}[htbp!]
\centering
  \subfloat[\centering ]{{\includegraphics[width=170pt,height=170pt]{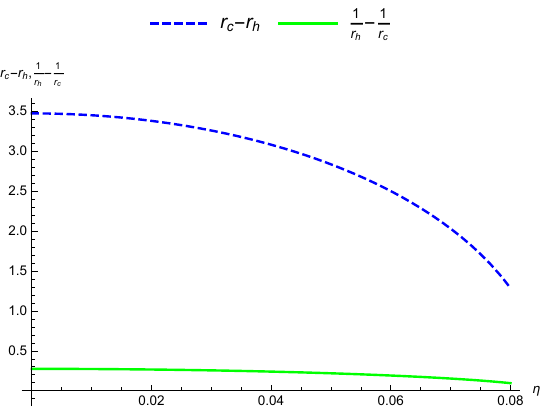}}\label{fig rcrce}}
  \qquad
   \subfloat[\centering ]{{\includegraphics[width=170pt,height=170pt]{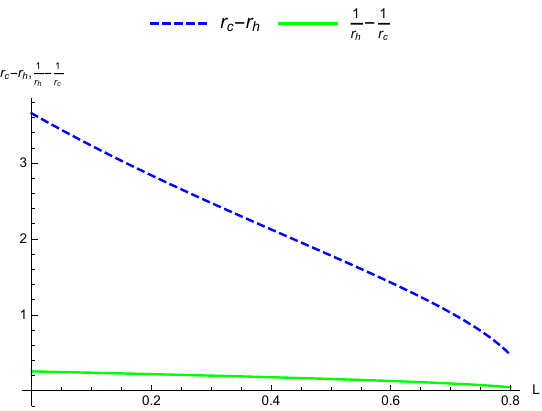}}\label{fig rcreL}}
   \caption{Graphs illustrating the behavior of $r_c-r_e$ and $\frac{1}{r_e}-\frac{1}{r_c}$ versus $\eta$ and $L$ respectively.}
   \label{fig rcrh}
\end{figure}
 From Fig. \ref{fig rcrh}, one can see that both $r_c-r_e$ and $\frac{1}{r_e}-\frac{1}{r_c}$ are strictly positive and decreasing with increasing
$\eta$ and $L$. This confirms that  the angle in RHS of \eqref{eqn egrey} is positive and it decreases  with decreasing the horizon distance.

\end{document}